\documentstyle [12pt,twoside, epsf]{article}
 \let\\=\cr
 
 \def\Chi{\hbox{\raise0.5ex\hbox{$\chi$}}}

\makeatletter
\long\def\M#1{\leavevmode\setbox\@tempboxa\hbox{#1}\@tempdima\fboxrule
     \advance\@tempdima \fboxsep \advance\@tempdima \dp\@tempboxa
    \hbox{\lower \@tempdima\hbox
   {\vbox{\hrule \@height \fboxrule
           \hbox{  \hskip\fboxsep
           \vbox{\vskip\fboxsep \box\@tempboxa\vskip\fboxsep}\hskip
                  \fboxsep\vrule \@width \fboxrule}%
                   }}}}

\makeatother

\def\picill#1by#2(#3)
{\vbox to #2
{\hrule width #1 height 0pt depth 0pt
\vfill\epsffile{#3}}}
 
\begin{document}

\date{}

\title{\Large\bf $q$ - Deformed Spin Networks, Knot Polynomials and Anyonic Topological Quantum Computation}
\author{Louis
H. Kauffman\\ Department of Mathematics, Statistics \\ and Computer Science (m/c
249)    \\ 851 South Morgan Street   \\ University of Illinois at Chicago\\
Chicago, Illinois 60607-7045\\ $<$kauffman@uic.edu$>$\\ and \\ Samuel J. Lomonaco
Jr. \\ Department of Computer Science and Electrical Engineering \\ University of
Maryland Baltimore County \\ 1000 Hilltop Circle, Baltimore, MD 21250\\
$<$lomonaco@umbc.edu$>$}

\maketitle

\thispagestyle{empty}

\subsection*{\centering Abstract}

{\em We review the $q$-deformed spin network approach to Topological Quantum Field Theory and apply these methods to
produce unitary representations of the braid groups that are dense in the unitary groups. Our methods are rooted in the bracket state sum model 
for the Jones polynomial. We give our results for a large class of representations based on values for the bracket polynomial that are roots of unity.
We make a separate and self-contained study of the quantum universal Fibonacci model in this framework. We apply our results to give quantum algorithms for
the computation of the colored Jones polynomials for knots and links, and the Witten-Reshetikhin-Turaev invariant of three manifolds.}

\section{Introduction}
This paper describes the background for topological quantum computing in terms of Temperley--Lieb recoupling theory.
This is a recoupling theory that generalizes standard angular momentum recoupling theory, generalizes the Penrose theory of spin networks and is inherently
topological.  Temperley--Lieb recoupling Theory is based on the bracket polynomial model \cite{KA87,KP} for the Jones polynomial. It is built in terms of
diagrammatic combinatorial topology. The same structure can be explained in terms of the $SU(2)_{q}$ quantum group, and has relationships with functional
integration and Witten's approach to topological quantum field theory. Nevertheless, the approach given here will be unrelentingly elementary. Elementary,
does not necessarily mean simple. In this case an architecture is built from simple beginnings and this archictecture and its recoupling language can be
applied to many things including, e.g.  colored Jones polynomials, Witten--Reshetikhin--Turaev invariants of three manifolds, topological quantum field theory
and quantum computing.
\bigbreak

In quantum computing, the application is most interesting because the simplest non-trivial example of this Temperley--Lieb recoupling Theory gives the
so-called Fibonacci model. The recoupling theory yields representations of the Artin Braid group into unitary groups
$U(n)$ where $n$ is a Fibonacci number. These representations are {\em dense} in the unitary group, and can be used to model quantum computation
universally in terms of representations of the braid group. Hence the term: topological quantum computation.
\bigbreak

In this paper, we outline the basics of the TL-Recoupling Theory, and show explicitly how the Fibonacci model arises from it.
The diagrammatic computations in the section 9 are completely self-contained and can be used by a reader who has just learned the bracket polynomial, and
wants to see how these dense unitary braid group representations arise from it. The outline of the parts of this paper is give below.
\bigbreak

\begin{enumerate}
\item Knots and Braids
\item Quantum Mechanics and Quantum Computation
\item $SU(2)$ Representations of the Artin Braid Group
\item The Bracket Polynomial and the Jones Polynomial
\item Quantum Topology, Cobordism Categories, Temperley-Lieb Algebra and Topological Quantum Field Theory
\item Braiding and Topological Quantum Field Theory
\item Spin Networks and Temperley-Lieb Recoupling Theory
\item Fibonacci Particles
\item The Fibonacci Recoupling Model
\item Quantum Computation of Colored Jones Polynomials and the Witten-Reshetikhin-Turaev Invariant
\end{enumerate}

We should point out that most of the results in this paper are either new, or are
new points of view on known results. The material on $SU(2)$ representations of the Artin braid group is new, and the relationship of this material to
the recoupling theory is new. See Theorem 1 in Section 3. The treatment of elementary cobordism categories is well-known, but new in the context of quantum
information theory. The reformulation of Temperley-Lieb recoupling theory for the purpose of producing unitary braid group representations is new for quantum
information  theory, and directly related to much of the recent work of Freedman and his collaborators. In Theorem 2 in Section 7 we give a general method
to obtain unitary representations at roots of unity using the recoupling theory. The treatment of the Fibonacci model in terms of two-strand  recoupling
theory is new and at the same time, the most elementary non-trivial example of the recoupling theory. The models in section 10 for quantum computation of
colored Jones polynomials and for quantum computation of the Witten-Reshetikhin-Turaev invariant are new. They take a
particularly simple aspect in this context.
\bigbreak

Here is a very condensed presentation of how unitary representations of the braid group are constructed via topological quantum field theoretic methods.
One has a mathematical particle with label $P$
that can interact with itself to produce either itself labeled $P$ or itself
with the null label $*.$ When $*$ interacts with $P$ the result is always $%
P. $ When $*$ interacts with $*$ the result is always $*.$ One considers
process spaces where a row of particles labeled $P$ can successively
interact subject to the restriction that the end result is $P.$ For example
the space $V[(ab)c]$ denotes the space of interactions of three particles
labeled $P.$ The particles are placed in the positions $a,b,c.$ Thus we
begin with $(PP)P.$ In a typical sequence of interactions, the first two $P$%
's interact to produce a $*,$ and the $*$ interacts with $P$ to produce $P.$ 
\[
(PP)P \longrightarrow (*)P \longrightarrow P. 
\]
\noindent In another possibility, the first two $P$'s interact to produce a $%
P,$ and the $P$ interacts with $P$ to produce $P.$ 
\[
(PP)P \longrightarrow (P)P \longrightarrow P. 
\]
It follows from this analysis that the space of linear combinations of
processes $V[(ab)c]$ is two dimensional. The two processes we have just
described can be taken to be the qubit basis for this space. One obtains
a representation of the three strand Artin braid group on $V[(ab)c]$ by
assigning appropriate phase changes to each of the generating processes. One
can think of these phases as corresponding to the interchange of the
particles labeled $a$ and $b$ in the association $(ab)c.$ The other operator
for this representation corresponds to the interchange of $b$ and $c.$ This
interchange is accomplished by a {\it unitary change of basis mapping} 
\[
F:V[(ab)c] \longrightarrow V[a(bc)]. 
\]
\noindent If 
\[
A:V[(ab)c] \longrightarrow V[(ba)c] 
\]
is the first braiding operator (corresponding to an interchange of the first
two particles in the association) then the second operator 
\[
B:V[(ab)c] \longrightarrow V[(ac)b] 
\]
is accomplished via the formula $B = F^{-1}RF$ where the $R$ in this formula
acts in the second vector space $V[a(bc)]$ to apply the phases for the
interchange of $b$ and $c.$ These issues are illustrated in Figure 1, where the parenthesization of the particles 
is indicated by circles and by also by trees. The trees can be taken to indicate patterns of particle interaction, where
two particles interact at the branch of a binary tree to produce the particle product at the root. See also Figure 28 for an illustration
of the braiding $B = F^{-1}RF$
\bigbreak

In this scheme, vector spaces corresponding to associated strings of
particle interactions are interrelated by {\it recoupling transformations}
that generalize the mapping $F$ indicated above. A full representation of
the Artin braid group on each space is defined in terms of the local
interchange phase gates and the recoupling transformations. These gates and
transformations have to satisfy a number of identities in order to produce a
well-defined representation of the braid group. These identities were
discovered originally in relation to topological quantum field theory. In
our approach the structure of phase gates and recoupling
transformations arise naturally from the structure of the bracket model for
the Jones polynomial. Thus we obtain a knot-theoretic basis for topological
quantum computing. \bigbreak

$$ \picill3inby4.5in(F1)  $$
\begin{center}
{\bf Figure 1  - Braiding Anyons. }
\end{center}

Aspects of the quantum Hall effect are related to topological quantum field theory
\cite{Wilczek,Fradkin,B1,B2}, where, in two dimensional space, the braiding of  quasi-particles or collective excitations leads to non-trival
representations of the Artin braid group. Such particles are called {\it Anyons}.  It is hoped that the mathematics we explain here will 
form the bridge between theoretical models of anyons and their applications to quantum computing.
\bigbreak

\noindent {\bf Acknowledgement.} The first author thanks the
National Science Foundation for support of this research under NSF Grant
DMS-0245588. Much of this effort was sponsored by the Defense
Advanced Research Projects Agency (DARPA) and Air Force Research Laboratory, Air
Force Materiel Command, USAF, under agreement F30602-01-2-05022. 
The U.S. Government is authorized to reproduce and distribute reprints
for Government purposes notwithstanding any copyright annotations thereon. The
views and conclusions contained herein are those of the authors and should not be
interpreted as necessarily representing the official policies or endorsements,
either expressed or implied, of the Defense Advanced Research Projects Agency,
the Air Force Research Laboratory, or the U.S. Government. (Copyright 2006.)  
It gives the authors pleasure to thank the Newton Institute in Cambridge England and ISI in Torino, Italy for their hospitality during 
the inception of this research and to thank Hilary Carteret for useful conversations.
\bigbreak

\section{Knots and Braids}
The purpose of this section is to give a quick introduction to the diagrammatic theory of knots,
links and braids. A {\it knot} is an embedding of a circle in three-dimensional space, taken up to ambient 
isotopy. The problem of deciding whether two knots are isotopic is an example of a {\it placement problem}, a problem of studying the topological forms that
can be made by placing one space inside another. In the case of knot theory we consider the placements of a circle inside three dimensional space.
There are many applications of the theory of knots. Topology is a background for the physical structure of real
knots made from rope of cable. As a result, the field of practical knot tying is a field of applied topology that existed well before the 
mathematical discipline of topology arose. Then again long molecules such as rubber molecules and DNA molecules can be knotted and linked. There have been
a number of intense applications of knot theory to the study of $DNA$ \cite{Sumners1} and to polymer physics \cite{KA}. Knot theory is closely related to
theoretical physics as well with applications in quantum gravity \cite{Smolin,CRLS,KaufLiko} and many applications of ideas in physics to the topological
structure of  knots themselves \cite{KP}. 
\bigbreak

{\it Quantum topology} is the study and invention of topological invariants via the use
of analogies and techniques from mathematical physics. Many invariants such as the Jones polynomial are 
constructed via partition functions and generalized quantum amplitudes. As a result, one expects to see relationships between knot theory
and physics. In this paper we will study how knot theory can be used to produce unitary representations of the braid group. Such representations
can play a fundamental role in quantum computing.
\bigbreak

$$ \picill5inby2in(F2)  $$
\begin{center}
{\bf Figure 2  - A knot diagram. }
\end{center}

$$ \picill5inby2.5in(F3)  $$
\begin{center}
{\bf Figure 3  - The Reidemeister Moves. }
\end{center}

\noindent That is, two knots are regarded as equivalent if one embedding can be obtained from the other
through a continuous family of embeddings of circles in three-space. A {\it link} is an embedding of a disjoiint
collection of circles, taken up to ambient isotopy. Figure 2 illustrates a diagram for a knot. The diagram is regarded
both as a schematic picture of the knot, and as a plane graph with extra structure at the nodes (indicating how the curve of 
the knot passes over or under itself by standard pictorial conventions).

$$ \picill5inby4in(F4)  $$
\begin{center}
{\bf Figure 4  - Braid Generators. }
\end{center}

Ambient isotopy is mathematically the same as the equivalence relation generated on diagrams by the {\it Reidemeister moves}. These moves are
illustrated in Figure 3. Each move is performed on a local part of the diagram that is topologically identical to the part of the diagram illustrated
in this figure (these figures are representative examples of the types of Reidemeister moves) without changing the rest of the diagram. The Reidemeister
moves are useful in doing combinatorial topology with knots and links, notably in working out the behaviour of knot invariants. A {\it knot invariant}
is a function defined from knots and links to some other mathematical object (such as groups or polynomials or numbers) such that equivalent diagrams are
mapped to equivalent objects (isomorphic groups, identical polynomials, identical numbers). The Reidemeister moves are of great use for analyzing the 
structure of knot invariants and they are closely related to the {\it Artin Braid Group}, which we discuss below.

$$ \picill5inby3in(F5)  $$
\begin{center}
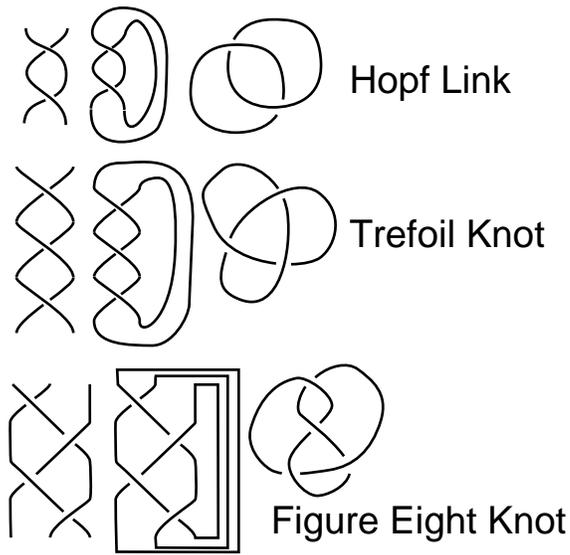
{\bf Figure 5  - Closing Braids to form knots and links. }
\end{center}

$$ \picill5inby2.5in(F6)  $$
\begin{center}
{\bf Figure 6  - Borromean Rings as a Braid Closure. }
\end{center}

Another significant structure related to knots and links is the Artin Braid Group. A {\it braid} is an embedding of a collection of strands that have 
their ends top and bottom row points in two rows of points that are set one above the other with respect to a choice of vertical. The strands are not
individually knotted and they are disjoint from one another. See Figures 4, 5 and 6 for illustrations of braids and moves on braids. Braids can be 
multiplied by attaching the bottom row of one braid to the top row of the other braid. Taken up to ambient isotopy, fixing the endpoints, the braids form
a group under this notion of multiplication. In Figure 4 we illustrate the form of the basic generators of the braid group, and the form of the
relations among these generators. Figure 5 illustrates how to close a braid by attaching the top strands to the bottom strands by a collection of 
parallel arcs. A key theorem of Alexander states that every knot or link can be represented as a closed braid. Thus the theory of braids is critical to the 
theory of knots and links. Figure 6 illustrates the famous Borromean Rings (a link of three unknotted loops such that any two of the loops are unlinked)
as the closure of a braid.
\bigbreak

Let $B_{n}$ denote the Artin braid group on $n$ strands.
We recall here that $B_{n}$ is generated by elementary braids $\{ s_{1}, \cdots ,s_{n-1} \}$
with relations 

\begin{enumerate}
\item $s_{i} s_{j} = s_{j} s_{i}$ for $|i-j| > 1$, 
\item $s_{i} s_{i+1} s_{i} = s_{i+1} s_{i} s_{i+1}$ for $i= 1, \cdots n-2.$
\end{enumerate}

\noindent See Figure 4 for an illustration of the elementary braids and their relations. Note that the braid group has a diagrammatic
topological interpretation, where a braid is an intertwining of strands that lead from one set of $n$ points to another set of $n$ points.
The braid generators $s_i$ are represented by diagrams where the $i$-th and $(i + 1)$-th strands wind around one another by a single 
half-twist (the sense of this turn is shown in Figure 4) and all other strands drop straight to the bottom. Braids are diagrammed
vertically as in Figure 4, and the products are taken in order from top to bottom. The product of two braid diagrams is accomplished by
adjoining the top strands of one braid to the bottom strands of the other braid. 
\bigbreak 

In Figure 4 we have restricted the illustration to the
four-stranded braid group $B_4.$ In that figure the three braid generators of $B_4$ are shown, and then the inverse of the
first generator is drawn. Following this, one sees the identities $s_{1} s_{1}^{-1} = 1$ 
(where the identity element in $B_{4}$ consists in  four vertical strands), 
$s_{1} s_{2} s_{1} = s_{2} s_{1}s_{2},$ and finally
$s_1 s_3 = s_3 s_1.$ 
\bigbreak

Braids are a key structure in mathematics. It is not just that they are a collection of groups with a vivid topological interpretation.
From the algebraic point of view the braid groups $B_{n}$ are important extensions of the symmetric groups $S_{n}.$ Recall that the 
symmetric group $S_{n}$ of all permutations of $n$ distinct objects has presentation as shown below.
\begin{enumerate}
\item $s_{i}^{2} = 1$ for  $i= 1, \cdots n-1,$
\item $s_{i} s_{j} = s_{j} s_{i}$ for $|i-j| > 1$, 
\item $s_{i} s_{i+1} s_{i} = s_{i+1} s_{i} s_{i+1}$ for $i= 1, \cdots n-2.$
\end{enumerate}
Thus $S_{n}$ is obtained from $B_{n}$ by setting the square of each braiding generator equal to one. We have an exact sequence of groups
$${1} \longrightarrow B_{n} \longrightarrow S_{n} \longrightarrow {1}$$ exhibiting the Artin Braid group as an extension of the symmetric group.
\bigbreak

In the next sections we shall show how representations of the Artin Braid group are rich enough to provide a dense set of transformations in the 
unitary groups. Thus the braid groups are, {\it in principle} fundamental to quantum computation and quantum information theory.
\bigbreak

\section{Quantum Mechanics and Quantum Computation}
We shall quickly
indicate the basic principles of quantum mechanics.  The quantum information context 
encapsulates a concise model of quantum theory:
\bigbreak

{\em The initial state of a quantum process is a vector $|v \rangle$ in a complex vector space $H.$
Measurement returns basis elements $\beta$ of $H$ with probability 

$$|\langle \beta \,|v \rangle |^{2}/\langle v \,|v \rangle$$

\noindent where $\langle v \,|w \rangle = v^{\dagger}w$ with $v^{\dagger}$ the conjugate transpose of $v.$
A physical process occurs in steps $|v\rangle \longrightarrow U\,|v \rangle = |Uv \rangle $ where $U$ is a unitary linear transformation.
\bigbreak

Note that since $\langle Uv \,|Uw \rangle = \langle v \,|U^{\dagger}U |w \rangle = \langle v \,|w \rangle = $ when $U$ is unitary, it follows that probability
is preserved in the  course of a quantum process.  }
\bigbreak

One of the details for any specific quantum problem is the nature of the unitary 
evolution.  This is specified by knowing appropriate information about the classical physics that 
supports the phenomena. This information is used to choose an appropriate Hamiltonian through which the 
unitary operator is constructed via a correspondence principle that replaces classical variables with appropriate quantum
operators. (In the path integral approach one needs a Langrangian to construct the action on which the path
integral is based.) One needs to know certain aspects of classical physics to 
solve any specific quantum problem.  
\bigbreak

A key concept in the quantum information viewpoint is the notion of the superposition of states.
If a quantum system has two  distinct states $|v \rangle$ and $|w \rangle,$ then it has infinitely many states of the form
$a|v \rangle + b|w \rangle$ where $a$ and $b$ are complex numbers taken up to a common multiple. States are ``really" 
in the projective space associated with $H.$ There is only one superposition of a single state $|v \rangle$ with 
itself. 
\bigbreak

Dirac \cite{D} introduced the ``bra -(c)-ket" notation $\langle A\,|B \rangle = A^{\dagger}B$ for the inner product of complex vectors $A,B \in H$.
He also separated the parts of the bracket into the {\em bra} $<A\,|$ and the {\em ket} $|B \rangle.$ Thus

$$\langle A\,|B \rangle = \langle A\,|\,\,|B \rangle$$

\noindent In this interpretation,
the ket $|B \rangle$ is identified with the vector $B \in H$, while the bra $<A\,|$ is regarded as the element dual to $A$ in the 
dual space $H^*$. The dual element to $A$ corresponds to the conjugate transpose $A^{\dagger}$ of the vector $A$, and the inner product is 
expressed in conventional language by the matrix product $A^{\dagger}B$ (which is a scalar since $B$ is a column vector). Having separated the bra and the ket, Dirac can write the
``ket-bra"  $|A \rangle \langle B\,| = AB^{\dagger}.$ In conventional notation, the ket-bra is a matrix, not a scalar, and we have the following formula for
the  square of $P = |A \rangle \langle B\,|:$

$$P^{2} =  |A \rangle \langle B\,| |A \rangle \langle B\,| = A(B^{\dagger}A)B^{\dagger} = (B^{\dagger}A)AB^{\dagger} = \langle B\,|A \rangle P.$$

\noindent The standard example is a ket-bra $P = |A\,\rangle \langle A|$ where $\langle A\,|A \rangle =1$ so that $P^2 = P.$  Then $P$ is a projection
matrix,  projecting to the subspace of $H$ that is spanned by the vector $|A \rangle$. In fact, for any vector $|B \rangle$ we have 

$$P|B \rangle = |A \rangle \langle A\,|\,|B \rangle =  |A \rangle \langle A\,|B \rangle = \langle A\,|B \rangle |A \rangle .$$

\noindent If $\{|C_{1} \rangle, |C_{2} \rangle , \cdots |C_{n} \rangle \}$ is an orthonormal basis for $H$, and $$P_{i} = |C_{i} \,\rangle \langle C_{i}|,$$
\noindent then for any vector $|A \rangle $ we have

$$|A \rangle = \langle C_{1}\,|A \rangle |C_{1} \rangle + \cdots + \langle C_{n}\,|A \rangle |C_{n} \rangle .$$

\noindent Hence 

$$\langle B\,|A \rangle = \langle B\,|C_{1} \rangle \langle C_{1}\,|A \rangle + \cdots + \langle B\,|C_{n} \rangle \langle C_{n}\,|A \rangle $$

One wants the probability of starting in state $|A \rangle $ and ending in state $|B \rangle .$ The 
probability for this event is equal to $|\langle B\,|A \rangle |^{2}$. This can be refined if we have more knowledge. 
If the intermediate states $|C_{i} \rangle $ are a complete set of orthonormal alternatives then we
can assume that 
$\langle C_{i}\,|C_{i} \rangle  = 1$ for each $i$ and that $\Sigma_{i} |C_{i} \rangle \langle C_{i}| = 1.$  This identity now corresponds to the fact that
$1$ is the sum of the probabilities of an arbitrary state being projected into one of these intermediate states.
\bigbreak

If there are intermediate states between the intermediate states this formulation can be continued
until one is summing over all possible paths from $A$ to $B.$ This becomes the path integral expression 
for the amplitude $\langle B|A \rangle .$
\bigbreak

\subsection{What is a Quantum Computer?}

A {\it quantum computer} is, abstractly, a composition $U$ of unitary transformations, together with an initial state and a choice of measurement
basis. One runs the computer by repeatedly initializing it, and then measuring the result of applying the unitary transformation $U$ to the initial state.
The results of these measurements are then analyzed for the desired information that the computer was set to determine. The key to using the computer
is the design of the initial state and the design of the composition of unitary transformations. The reader should consult \cite{N} for more specific
examples of quantum algorithms. 
\bigbreak

Let $H$ be a given finite dimensional vector space over the complex numbers $C.$ Let $\{ W_{0}, W_{1},..., W_{n} \}$ be an
orthonormal basis for $H$ so that with $|i \rangle := |W_{i} \rangle $ denoting $W_{i}$ and $\langle i|$ denoting the conjugate transpose of $|i \rangle $,
we have
$$\langle i|j \rangle = \delta_{ij}$$
\noindent where $\delta_{ij}$ denotes the Kronecker delta (equal to one when its indices are equal to one another, and equal
to zero otherwise). Given a vector $v$ in $H$ let $|v|^{2} := \langle v|v \rangle .$ Note that $\langle i|v$ is the $i$-th coordinate of $v.$ 
\vspace{3mm}

\noindent An {\em measurement of $v$} returns one of the coordinates $|i \rangle $
of $v$ with probability $|\langle i|v|^{2}.$ This model of measurement is a simple instance of the situation with a quantum
mechanical system that is in a mixed state until it is measured. The result of measurement is to put the system into one of
the basis states. 
\vspace{3mm}

When the dimension of the space $H$ is two ($n=1$), a vector in the space is called a {\em qubit}. A qubit represents one
quantum of binary information. On meausurement, one obtains either the ket $|0 \rangle $ or the ket $|1 \rangle $. This constitutes the 
binary distinction that is inherent in a qubit.  Note however that the information obtained is probabilistic.  If the qubit is
$$| \psi \rangle = \alpha |0 \rangle + \beta \ |1 \rangle ,$$ \noindent then the ket $|0 \rangle $ is observed with probability $|\alpha|^{2}$, and the ket
$|1 \rangle $ is observed with probability $|\beta|^{2}.$  In speaking of an idealized quantum computer, we do not specify the nature
of measurement process beyond these probability postulates.
\vspace{3mm}
 
In the case of general dimension $n$ of the space $H$, we will call the vectors in $H$
{\em qudits}. It is quite common to use spaces $H$ that are tensor products of two-dimensional spaces (so that all computations 
are expressed in terms of qubits) but this is not necessary in principle. One can start with a given space, and later work out
factorizations into qubit transformations.
\vspace{3mm}

A {\em quantum computation} consists in the application of a unitary
transformation $U$ to an initial qunit $\psi = a_{0}|0 \rangle + ... + a_{n}|n \rangle $  with $|\psi|^{2}=1$, plus a measurement of
$U\psi.$ A measurement of $U\psi$ returns the ket $|i \rangle $ with probability $|\langle i|U\psi|^2$. In particular, if we start the computer
in the state $|i \rangle $, then the probability that it will return the state $|j \rangle $ is $|\langle j|U|i \rangle |^{2}.$

\vspace{3mm} It is the necessity for writing a given computation in terms of unitary transformations, and the probabilistic
nature of the result that characterizes quantum computation. Such computation could be carried out by an idealized quantum
mechanical system. It is hoped that such systems can be physically realized. 
\vspace{3mm}

 \subsection{Universal Gates}
A {\em two-qubit gate} $G$ is a unitary  linear mapping $G:V \otimes V \longrightarrow V$ where $V$ is a two complex dimensional
vector space. We say that the gate $G$ is {\em universal for quantum computation} (or just {\em universal}) if $G$ together with 
local unitary transformations (unitary transformations from $V$ to $V$) generates all unitary transformations of the complex vector
space of dimension $2^{n}$ to itself. It is well-known \cite{N} that $CNOT$ is a universal gate. (On the standard basis,
$CNOT$ is the identity when the first qubit is $0$, and it flips the second qbit, leaving the first alone, when the first qubit is $1.$)
\bigbreak

\noindent A gate $G$, as above, is said to be {\em entangling} if there is a vector  
$$| \alpha \beta \rangle = | \alpha \rangle \otimes | \beta \rangle \in V \otimes V$$ such that 
$G | \alpha \beta \rangle$ is not decomposable as a tensor product of two qubits. Under these circumstances, one says that 
$G | \alpha \beta \rangle$ is {\em entangled}.
\bigbreak

\noindent In \cite{BB},  the Brylinskis
give a general criterion of $G$ to be universal. They prove that {\em a two-qubit gate $G$ is universal if and only if it is
entangling.} 
\bigbreak

\noindent {\bf Remark.} A two-qubit pure state $$|\phi \rangle = a|00 \rangle + b|01 \rangle + c|10 \rangle + d|11 \rangle$$
is entangled exactly when $(ad-bc) \ne 0.$ It is easy to use this fact to check when a specific matrix is, or is not, entangling.
\bigbreak

\noindent {\bf Remark.} There are many gates other than $CNOT$ that can be used as universal gates in the presence of local unitary 
transformations. Some of these are themselves topological (unitary solutions to the Yang-Baxter equation, see \cite{BG,Yong}) and themselves generate
representations of the Artin Braid Group. Replacing $CNOT$ by a solution to the Yang-Baxter equation does not place the local unitary transformations as
part of the corresponding representation of the braid group. Thus such substitutions connote only a partial solution to creating topological 
quantum computation. In this paper we are concerned with braid group representations that include all aspects of the unitary
group. Accordingly, in the next section we shall first examine, how the braid group on three strands can be represented as local unitary
transformations.
\bigbreak

\section{$SU(2)$ Representations of the Artin Braid Group}
The purpose of this section is to determine all the representations of the three strand Artin braid group $B_{3}$ to the special unitary group $SU(2)$ and
concomitantly to the unitary group $U(2).$ One regards the groups $SU(2)$ and $U(2)$ as acting on a single qubit, and so $U(2)$ is usually regarded as the
group of local unitary transformations in a quantum information setting. If one is looking for a coherent way to represent all unitary transformations by
way of braids, then $U(2)$ is the place to start. Here we will show that there are many representations of the three-strand braid group
that generate a dense subset of $U(2).$ Thus it is a fact that local unitary transformations can be "generated by braids" in many ways.
\bigbreak

We begin with the structure of $SU(2).$ A matrix in $SU(2)$ has the form 
$$ M = 
\left( \begin{array}{cc}
z & w \\
-\bar{w} & \bar{z} \\
\end{array} \right),$$ where $z$ and $w$ are complex numbers, and $\bar{z}$ denotes the complex conjugate of $z.$ 
To be in $SU(2)$ it is required that $Det(M)=1$ and that $M^{\dagger} = M^{-1}$ where $Det$ denotes determinant, and $M^{\dagger}$ is the conjugate transpose of $M.$
Thus if
$z = a + bi$ and $w = c + di$ where $a,b,c,d$ are real numbers, and $i^2 = -1,$ then 
$$ M = 
\left( \begin{array}{cc}
a + bi & c + di \\
-c + di & a - bi \\
\end{array} \right)$$  with $a^2 + b^2 + c^2 + d^2 = 1.$ It is convenient to write
$$M =
a\left( \begin{array}{cc}
1 & 0 \\
0 & 1 \\
\end{array} \right) +
b\left( \begin{array}{cc}
i & 0\\
0 & -i \\
\end{array} \right) +
c\left( \begin{array}{cc}
0  & 1 \\
-1  & 0\\
\end{array} \right) +
d\left( \begin{array}{cc}
0 & i \\
i & 0 \\
\end{array} \right),$$ and to abbreviate this decomposition as
$$M = a + bi +cj + dk$$
where 
$$ 1 \equiv
\left( \begin{array}{cc}
1 & 0 \\
0 & 1 \\
\end{array} \right),
i \equiv
\left( \begin{array}{cc}
i & 0\\
0 & -i \\
\end{array} \right),
j \equiv,
\left( \begin{array}{cc}
0  & 1 \\
-1  & 0\\
\end{array} \right),
k \equiv
\left( \begin{array}{cc}
0 & i \\
i & 0 \\
\end{array} \right)$$ so that 
$$i^2 = j^2 = k^2 = ijk = -1$$ and 
$$ij = k, jk=i, ki = j$$
$$ji = -k, kj = -i, ik = -j.$$
The algebra of $1,i,j,k$ is called the {\it quaternions} after William Rowan Hamilton who discovered this algebra prior to the discovery of 
matrix algebra. Thus the unit quaternions are identified with $SU(2)$ in this way. We shall use this identification, and some facts about 
the quaternions to find the $SU(2)$ representations of braiding. First we recall some facts about the quaternions.

\begin{enumerate}
\item Note that if $q = a + bi +cj + dk$ (as above), then $q^{\dagger} = a - bi - cj - dk$ so that $qq^{\dagger} = a^2 + b^2 + c^2 + d^2 = 1.$
\item A general quaternion has the form $ q = a + bi + cj + dk$ where the value of $qq^{\dagger} = a^2 + b^2 + c^2 + d^2,$ is not fixed to unity.
The {\it length} of $q$ is by definition $\sqrt{qq^{\dagger}}.$
\item A quaternion of the form $ri + sj + tk$ for real numbers $r,s,t$ is said to be a {\it pure} quaternion. We identify the set of pure
quaternions with the vector space of triples $(r,s,t)$ of real numbers $R^{3}.$
\item Thus a general quaternion has the form $q = a + bu$ where $u$ is a pure quaternion of unit length and $a$ and $b$ are arbitrary real numbers.
A unit quaternion (element of $SU(2)$) has the addition property that $a^2 + b^2 = 1.$
\item If $u$ is a pure unit length quaternion, then $u^2 = -1.$ Note that the set of pure unit quaternions forms the two-dimensional sphere
$S^{2} = \{ (r,s,t) | r^2 + s^2 + t^2 = 1 \}$ in $R^{3}.$
\item If $u, v$ are pure quaternions, then $$uv = -u \cdot v + u \times v$$ whre $u \cdot v$ is the dot product of the vectors $u$ and $v,$ and 
$u \times v$ is the vector cross product of $u$ and $v.$ In fact, one can take the definition of quaternion multiplication as
$$(a + bu)(c + dv) = ac + bc(u) + ad(v) + bd(-u \cdot v + u \times v),$$ and all the above properties are consequences of this
definition. Note that quaternion multiplication is associative.
\item Let $g = a + bu$ be a unit length quaternion so that $u^2 = -1$ and $a = cos(\theta/2), b=sin(\theta/2)$ for a chosen angle $\theta.$
Define $\phi_{g}:R^{3} \longrightarrow R^{3}$ by the equation $\phi_{g}(P) = gPg^{\dagger},$ for $P$ any point in $R^{3},$ regarded as a pure quaternion.
Then $\phi_{g}$ is an orientation preserving rotation of $R^{3}$ (hence an element of the rotation group $SO(3)$). Specifically, $\phi_{g}$ is a rotation
about the  axis $u$ by the angle $\theta.$ The mapping $$\phi:SU(2) \longrightarrow SO(3)$$ is a two-to-one surjective map from the special unitary group to
the rotation group. In quaternionic form, this result was proved by Hamilton and by Rodrigues in the middle of the nineteeth century.
The specific formula for $\phi_{g}(P)$ as shown below:
$$\phi_{g}(P) = gPg^{-1} = (a^2 - b^2)P + 2ab (P \times u) + 2(P \cdot u)b^{2}u.$$
\end{enumerate}

We want a representation of the three-strand braid group in $SU(2).$ This means that we want a homomorphism $\rho: B_{3} \longrightarrow SU(2),$ and hence
we want elements $g = \rho(s_{1})$ and $h= \rho(s_{2})$ in $SU(2)$ representing the braid group generators $s_{1}$ and $s_{2}.$ Since $s_{1}s_{2}s_{1} =
s_{2}s_{1}s_{2}$ is the generating relation for $B_{3},$ the only requirement on $g$ and $h$ is that $ghg = hgh.$ We rewrite this relation as
$h^{-1}gh = ghg^{-1},$ and analyze its meaning in the unit quaternions.
\bigbreak

Suppose that $g = a + bu$ and $h=c + dv$ where $u$ and $v$ are unit pure quaternions so that $a^2 + b^2 = 1$ and $c^2 + d^2 = 1.$
then $ghg^{-1} = c +d\phi_{g}(v)$ and $h^{-1}gh = a + b\phi_{h^{-1}}(u).$ Thus it follows from the braiding relation that 
$a=c,$ $b= \pm d,$ and that $\phi_{g}(v) = \pm \phi_{h^{-1}}(u).$  However, in the case where there is a minus sign we have
$g = a + bu$ and $h = a - bv = a + b(-v).$ Thus we can now prove the following Theorem. 
\bigbreak

\noindent {\bf Theorem 1.} {\it If $g = a + bu$ and $h=c + dv$ are pure unit quaternions,then, without loss of generality, the braid relation $ghg=hgh$ is
true if and only if
$h = a + bv,$ and $\phi_{g}(v) = \phi_{h^{-1}}(u).$ Furthermore, given that $g = a +bu$ and $h = a +bv,$ the condition $\phi_{g}(v) = \phi_{h^{-1}}(u)$
is satisfied if and only if $u \cdot v = \frac{a^2 - b^2}{2 b^2}$ when $u \ne v.$ If $u = v$ then then $g = h$ and the braid relation is trivially
satisfied.}
\bigbreak

\noindent {\bf Proof.} We have proved the first sentence of the Theorem in the discussion prior to its statement. Therefore assume that
$g = a +bu, h = a +bv,$ and $\phi_{g}(v) = \phi_{h^{-1}}(u).$ 
We have already stated the formula for $\phi_{g}(v)$ in the discussion about quaternions:
$$\phi_{g}(v) = gvg^{-1} = (a^2 - b^2)v + 2ab (v \times u) + 2(v \cdot u)b^{2}u.$$ By the same token, we have
$$\phi_{h^{-1}}(u) = h^{-1}uh = (a^2 - b^2)u + 2ab (u \times -v) + 2(u \cdot (-v))b^{2}(-v)$$
$$= (a^2 - b^2)u + 2ab (v \times u) + 2(v \cdot u)b^{2}(v).$$ Hence we require that
$$(a^2 - b^2)v + 2(v \cdot u)b^{2}u = (a^2 - b^2)u + 2(v \cdot u)b^{2}(v).$$ This equation is equivalent to
$$2(u \cdot v)b^{2} (u - v) = (a^2 - b^2)(u - v).$$
If $u \ne v,$ then this implies that $$u \cdot v = \frac{a^2 - b^2}{2 b^2}.$$
This completes the proof of the Theorem.
$\hfill \Box$
\bigbreak

\noindent{\bf An Example.} Let
$$g = e^{i\theta} = a + bi$$ where $a = cos(\theta)$ and $b = sin(\theta).$
Let $$h = a + b[(c^2 - s^2)i + 2csk]$$ where $c^2 + s^2 = 1$ and $c^2 - s^2 = \frac{a^2 - b^2}{2b^2}.$ Then we can reexpress $g$ and $h$ in matrix form
as the matrices $G$ and $H.$ Instead of writing the explicit form of $H,$ we write $H = FGF^{\dagger}$ where $F$ is an element of $SU(2)$ as shown below.

$$G =
\left( \begin{array}{cc}
e^{i\theta} & 0 \\
0 & e^{-i\theta} \\
\end{array} \right)$$

$$F =
\left( \begin{array}{cc}
ic & is \\
is & -ic \\
\end{array} \right)$$
This representation of braiding where one generator $G$ is a simple matrix of phases, while the other generator $H = FGF^{\dagger}$ is derived from $G$ by
conjugation by a unitary matrix, has the possibility for generalization to representations of braid groups (on greater than three strands) to $SU(n)$ or
$U(n)$ for 
$n$ greater than $2.$ In fact we shall see just such representations constructed later in this paper, by using a version of topological quantum field theory.
The simplest example is given by 
$$g = e^{7 \pi i/10}$$
$$f = i \tau  + k \sqrt{\tau}$$
$$h = f r f^{-1}$$
where $\tau^{2} + \tau = 1.$
Then $g$ and $h$ satisfy $ghg=hgh$ and generate a representation of the three-strand braid group that is dense in $SU(2).$ We shall call this the 
{\it Fibonacci} representation of $B_{3}$ to $SU(2).$
\bigbreak

\noindent {\bf Density.} Consider representations of $B_{3}$ into $SU(2)$ produced by the method of this section. That is consider the subgroup $SU[G,H]$ of 
$SU(2)$ generated by a pair of elements $\{g,h \}$ such that $ghg=hgh.$ We wish to understand when such a representation will be dense in $SU(2).$
We need the following lemma.
\bigbreak

\noindent {\bf Lemma.} {\it $e^{ai} e^{bj} e^{ci} = cos(b) e^{i(a +c)} + sin(b) e^{i(a-c)} j.$ Hence any element of $SU(2)$ can be written in the form
$e^{ai} e^{bj} e^{ci}.$ for appropriate choices of angles $a,b,c.$ In fact, if $u$ and $v$ are linearly independent unit vectors in $R^{3},$ then
any element of $SU(2)$ can be written in the form $$e^{au} e^{bv} e^{cu}$$ for appropriate choices of the real numbers $a,b,c.$}
\bigbreak

\noindent {\bf Proof.}
It is easy to check that 
$$e^{ai} e^{bj} e^{ci} = cos(b)e^{i(a + c)} + sin(b)e^{i(a - c)} j.$$
This completes the verification of the identity in the statement of the Lemma.
\bigbreak

\noindent Let $v$ be any unit direction in $R^{3}$ and $\lambda$ an arbitrary angle.  
We have $$e^{v\lambda} = cos(\lambda) + sin(\lambda)v,$$ and 
$$v = r + si + (p + qi)j$$ where $r^2 + s^2 + p^2 + q^2 = 1.$ So
$$e^{v\lambda} = cos(\lambda) + sin(\lambda)[r + si] + sin(\lambda)[p + qi]j$$
$$= [(cos(\lambda) + sin(\lambda)r) + sin(\lambda)s i] + [sin(\lambda)p + sin(\lambda)q i]j.$$
\bigbreak

\noindent By the identity just proved, we can choose angles $a,b,c$ so that
$$e^{v\lambda} = e^{ia}e^{jb}e^{ic}.$$ Hence
$$cos(b)e^{i(a + c)} = (cos(\lambda) + sin(\lambda)r) + sin(\lambda)s i$$
and
$$sin(b)e^{i(a - c)} = sin(\lambda)p + sin(\lambda)q i.$$
Suppose we keep $v$ fixed and vary $\lambda.$ Then the last equations show that this will result in a full variation of $b.$
\bigbreak

\noindent Now consider 
$$e^{ia'}e^{v\lambda}e^{ic'} = e^{ia'}e^{ia}e^{jb}e^{ic}e^{ib'} = e^{i(a' + a)}e^{jb}e^{i(c + c')}.$$
By the basic identity, this shows that any element of $SU(2)$ can be written in the form 
$$e^{ia'}e^{v\lambda}e^{ic'}.$$
Then, by applying a rotation, we finally conclude that if $u$ and $v$ are linearly independent unit vectors in $R^{3},$ then any element of
$SU(2)$ can be written in the form
$$e^{au} e^{bv} e^{cu}$$ for appropriate choices of the real numbers $a,b,c.$
$\hfill \Box$
\bigbreak

This Lemma can be used to verify density of a representation, by finding two elements $A$ and $B$ in the representation such that 
the powers of $A$ are dense in the rotations about its axis, and the powers of $B$ are dense in the rotations about its axis, and such that the 
axes of $A$ and $B$ are linearly independent in $R^{3}.$ Then by the Lemma the set of elements $A^{a+c}B^{b}A^{a-c}$ are dense in $SU(2).$ It follows
for example, that the Fibonacci representation described above is dense in $SU(2),$ and indeed the generic representation of $B_{3}$ into
$SU(2)$ will be dense in $SU(2).$  Our next task is to describe representations of the higher braid groups that will extend some of these unitary
repressentations of the three-strand braid group. For this we need more topology.
\bigbreak

\section{The Bracket Polynomial and the Jones Polynomial}
We now discuss the Jones polynomial. We shall construct the Jones polynomial by using the bracket state 
summation model \cite{KA87}. The bracket polynomial, invariant under Reidmeister moves II and III, can be normalized to give an invariant of all three
Reidemeister moves. This normalized invariant, with a change of variable, is the Jones polynomial
\cite{JO1,JO2}. The Jones polynomial was originally discovered by a different method than the one given here. 
\bigbreak 

The {\em bracket polynomial} , $<K> \, = \, <K>(A)$,  assigns to each unoriented link diagram $K$ a 
Laurent polynomial in the variable $A$, such that
   
\begin{enumerate}
\item If $K$ and $K'$ are regularly isotopic diagrams, then  $<K> \, = \, <K'>$.
  
\item If  $K \sqcup O$  denotes the disjoint union of $K$ with an extra unknotted and unlinked 
component $O$ (also called `loop' or `simple closed curve' or `Jordan curve'), then 

$$< K \sqcup O> \, = \delta<K>,$$ 
where  $$\delta = -A^{2} - A^{-2}.$$
  
\item $<K>$ satisfies the following formulas 

$$<\mbox{\large $\chi$}> \, = A <\mbox{\large $\asymp$}> + A^{-1} <)(>$$
$$<\overline{\mbox{\large $\chi$}}> \, = A^{-1} <\mbox{\large $\asymp$}> + A <)(>,$$
\end{enumerate}

\noindent where the small diagrams represent parts of larger diagrams that are identical except  at
the site indicated in the bracket. We take the convention that the letter chi, \mbox{\large $\chi$},
denotes a crossing where {\em the curved line is crossing over the straight
segment}. The barred letter denotes the switch of this crossing, where {\em the curved
line is undercrossing the straight segment}.  See Figure 7 for a graphic illustration of this relation, and an
indication of the convention for choosing the labels $A$ and $A^{-1}$ at a given crossing.

$$ \picill5inby3in(F7) $$
\begin{center} {\bf Figure 7 - Bracket Smoothings} 
\end{center}
\vspace{3mm}

\noindent It is easy to see that Properties $2$ and $3$ define the calculation of the bracket on
arbitrary link diagrams. The choices of coefficients ($A$ and $A^{-1}$) and the value of $\delta$
make the bracket invariant under the Reidemeister moves II and III. Thus
Property $1$ is a consequence of the other two properties. 
\bigbreak

In computing the bracket, one finds the following behaviour under Reidemeister move I: 
  $$<\mbox{\large $\gamma$}> = -A^{3}<\smile> \hspace {.5in}$$ and 
  $$<\overline{\mbox{\large $\gamma$}}> = -A^{-3}<\smile> \hspace {.5in}$$

\noindent where \mbox{\large $\gamma$}  denotes a curl of positive type as indicated in Figure 8, 
and  $\overline{\mbox{\large $\gamma$}}$ indicates a curl of negative type, as also seen in this
figure. The type of a curl is the sign of the crossing when we orient it locally. Our convention of
signs is also given in Figure 8. Note that the type of a curl  does not depend on the orientation
we choose.  The small arcs on the right hand side of these formulas indicate
the removal of the curl from the corresponding diagram.  

\bigbreak
  
\noindent The bracket is invariant under regular isotopy and can be  normalized to an invariant of
ambient isotopy by the definition  
$$f_{K}(A) = (-A^{3})^{-w(K)}<K>(A),$$ where we chose an orientation for $K$, and where $w(K)$ is 
the sum of the crossing signs  of the oriented link $K$. $w(K)$ is called the {\em writhe} of $K$. 
The convention for crossing signs is shown in  Figure 8.

$$ \picill4.5inby2in(F8) $$
\begin{center} {\bf Figure 8 - Crossing Signs and Curls} 
\end{center}
\vspace{3mm}

\noindent One useful consequence of these formulas is the following {\em switching formula}
$$A<\mbox{\large $\chi$}> - A^{-1} <\overline{\mbox{\large $\chi$}}> = (A^{2} - A^{-2})<\mbox{\large $\asymp$}>.$$ Note that 
in these conventions the $A$-smoothing of $\mbox{\large $\chi$}$ is $\mbox{\large $\asymp$},$ while the $A$-smoothing of
$\overline{\mbox{\large $\chi$}}$ is $)(.$ Properly interpreted, the switching formula above says that you can switch a crossing and 
smooth it either way and obtain a three diagram relation. This is useful since some computations will simplify quite quickly with the 
proper choices of switching and smoothing. Remember that it is necessary to keep track of the diagrams up to regular isotopy (the 
equivalence relation generated by the second and third Reidemeister moves). Here is an example. View Figure 9.

$$ \picill3inby2in(F9) $$
\begin{center} {\bf Figure 9 -- Trefoil and Two Relatives} \end{center}
\bigbreak

\noindent Figure 9 shows a trefoil diagram $K$, an unknot diagram $U$ and another unknot diagram $U'.$  Applying the switching formula,
we have $$A^{-1} <K> - A <U> = (A^{-2} - A^{2}) <U'>$$ and 
$<U>= -A^{3}$ and $<U'>=(-A^{-3})^2 = A^{-6}.$  Thus $$A^{-1} <K> - A(-A^{3}) = (A^{-2} - A^{2}) A^{-6}.$$ Hence
$$A^{-1} <K> = -A^4 + A^{-8} - A^{-4}.$$  Thus $$<K> = -A^{5} - A^{-3} + A^{-7}.$$ This is the bracket polynomial of the trefoil diagram $K.$
\bigbreak

\noindent Since the trefoil diagram $K$ has writhe $w(K) = 3,$ we have the normalized polynomial 
$$f_{K}(A) = (-A^{3})^{-3}<K> = -A^{-9}(-A^{5} - A^{-3} + A^{-7}) = A^{-4} + A^{-12} - A^{-16}.$$  
\bigbreak

The bracket model for the Jones polynomial is quite useful both theoretically and in terms
 of practical computations. One of the neatest applications is to simply compute, as we have done, $f_{K}(A)$ for the
trefoil knot $K$ and determine that  $f_{K}(A)$ is not equal to $f_{K}(A^{-1}) = f_{-K}(A).$  This
shows that the trefoil is not ambient isotopic to its mirror image, a fact that is much harder to
prove by classical methods.
\bigbreak

\noindent {\bf The State Summation.} In order to obtain a closed formula for the bracket, we now describe it as a state summation.
Let $K$ be any unoriented link diagram. Define a {\em state}, $S$, of $K$  to be a choice of
smoothing for each  crossing of $K.$ There are two choices for smoothing a given  crossing, and
thus there are $2^{N}$ states of a diagram with $N$ crossings.
 In a state we label each smoothing with $A$ or $A^{-1}$ according to the left-right convention 
discussed in Property $3$ (see Figure 7). The label is called a {\em vertex weight} of the state.
There are two evaluations related to a state. The first one is the product of the vertex weights,
denoted  

$$<K|S>.$$
The second evaluation is the number of loops in the state $S$, denoted  $$||S||.$$
  
\noindent Define the {\em state summation}, $<K>$, by the formula 

$$<K> \, = \sum_{S} <K|S>\delta^{||S||-1}.$$
It follows from this definition that $<K>$ satisfies the equations
  
$$<\mbox{\large $\chi$}> \, = A <\mbox{\large $\asymp$}> + A^{-1} <)(>,$$
$$<K \sqcup  O> \, = \delta<K>,$$
$$<O> \, =1.$$
  
\noindent The first equation expresses the fact that the entire set of states of a given diagram is
the union, with respect to a given crossing, of those states with an $A$-type smoothing and those
 with an $A^{-1}$-type smoothing at that crossing. The second and the third equation
are clear from the formula defining the state summation. Hence this state summation produces the
bracket polynomial as we have described it at the beginning of the  section. 

\bigbreak

\noindent {\bf Remark.} By a change of variables one obtains the original
Jones polynomial, $V_{K}(t),$  for oriented knots and links from the normalized bracket:

$$V_{K}(t) = f_{K}(t^{-\frac{1}{4}}).$$

\noindent {\bf Remark.} The bracket polynomial provides a connection between  knot theory and physics, in that the state summation 
expression for it exhibits it as a generalized partition function defined on the knot diagram. Partition functions
are ubiquitous in statistical mechanics, where they express the summation over all states of the physical system of 
probability weighting functions for the individual states. Such physical partition functions contain large amounts of 
information about the corresponding physical system. Some of this information is directly present in the properties of the 
function, such as the location of critical points and phase transition. Some of the information can be obtained by differentiating the 
partition function, or performing other mathematical operations on it. 
\bigbreak

There is much more in this connection with statistical mechanics in that the local weights in a partition function are often expressed in
terms of solutions to a matrix equation called the Yang-Baxter equation, that turns out to fit perfectly invariance under the third 
Reidemeister move. As a result, there are many ways to define partition functions of knot diagrams that give rise to invariants of knots and links.
The subject is intertwined with the algebraic structure of Hopf algebras and quantum groups, useful for producing systematic solutions to the Yang-Baxter
equation. In fact Hopf algebras are deeply connected with the problem of constructing invariants of three-dimensional manifolds in relation to
invariants of knots. We have chosen, in this survey paper, to not discuss the details of these approaches, but rather to proceed to Vassiliev invariants
and the relationships with Witten's functional integral. The
reader is referred to \cite{KA87,KA89,KL,Kauffman-Graph,KaufInter,KP,AW,JO1,JO2,KR,RT1,RT2,T,TV} for more information about relationships of knot theory with
statistical mechanics, Hopf algebras and quantum groups. For topology, the key point is that Lie algebras can be used to construct invariants of
knots and links. 
\bigbreak

\subsection{Quantum Computation of the Jones Polynomial}

Quantum algorithms for computing the Jones polynomial have been discussed elsewhere. See \cite{QCJP,BG,Ah1,QCJP2,Ah2,Wo}. Here, as an example, we give a
local unitary representation that can be used to compute the Jones polynomial for closures of 3-braids.   We analyse this representation by making
explicit how the bracket polynomial is computed from it, and showing how the quantum computation devolves to finding the trace of a unitary transformation.
\vspace{3mm}

The idea behind the construction of this representation depends upon the algebra generated by two single qubit density matrices
(ket-bras).
Let
$|v\rangle$ and 
$|w\rangle$ be two qubits in $V,$ a complex vector space of dimension two over the complex numbers. Let
$P = |v\rangle\langle v|$ and $Q=|w\rangle\langle w|$ be the corresponding ket-bras.  Note that
$$P^2 = |v|^{2}P,$$
$$Q^2 = |w|^{2}Q,$$
$$PQP = |\langle v|w \rangle|^{2}P,$$ 
$$QPQ= |\langle v|w\rangle|^{2}Q.$$
$P$ and $Q$ generate a representation of the Temperley-Lieb algebra (See Section 5 of the present paper). One can adjust parameters to make a representation
of the three-strand braid group in the form
$$s_{1} \longmapsto rP + sI,$$
$$s_{2} \longmapsto tQ + uI,$$
where $I$ is the identity mapping on $V$ and $r,s,t,u$ are suitably chosen scalars. In the following we use this method to adjust
such a representation so that it is unitary.  Note also that this is a local unitary representation of $B_{3}$ to $U(2).$ We leave it as an exersise for
the reader to verify that it fits into our general classification of such representations as given in section 3 of the present paper.
\bigbreak

The representation depends on two symmetric but non-unitary matrices $U_{1}$ and $U_{2}$ with

$$U_{1} =  \left[
\begin{array}{cc}
     d & 0  \\
     0 & 0
\end{array}
\right] = d|w \rangle \langle w|$$

\noindent and 

$$U_{2} =  \left[
\begin{array}{cc}
     d^{-1} & \sqrt{1-d^{-2}}  \\
         \sqrt{1-d^{-2}}  & d - d^{-1}
\end{array}
\right] = d | v \rangle \langle v |$$
where $w = (1,0),$ and $v = (d^{-1}, \sqrt{1 - d^{-2}}),$ assuming the entries of $v$ are real.
Note that $U_{1}^{2} = dU_{1}$ and $U_{2}^{2} = dU_{1}.$ Moreover, $U_{1}U_{2}U_{1} = U_{1}$ and $U_{2}U_{1}U_{2} = U_{1}.$
 This is an example of a specific
representation of the Temperley-Lieb algebra \cite{KA87, QCJP}.
\noindent The desired representation of the Artin braid group is given on the two braid generators for the three strand braid group by the
equations:

$$\Phi(s_{1})= AI + A^{-1}U_{1},$$
$$\Phi(s_{2})= AI + A^{-1}U_{2}.$$
Here $I$ denotes the $2 \times 2$ identity matrix.

\noindent For any $A$ with $d = -A^{2}-A^{-2}$ these formulas define a representation of the braid group. With 
$A=e^{i\theta}$, we have $d = -2cos(2\theta)$. We find a specific range of
angles $|\theta| \leq \pi/6$ and $|\theta - \pi| \leq \pi/6$ {\it that give unitary representations of
the three-strand braid group.} Thus a specialization of a more general represention of the braid group gives rise to a continuous family
of unitary representations of the braid group.
\vspace{3mm}

Note that the traces of these matrices are given by the formulas $tr(U_{1})=tr(U_{2})= d$ while $tr(U_{1}U_{2}) = tr(U_{2}U_{1}) =1.$
If $b$ is any braid, let $I(b)$ denote the sum of the exponents in the braid word that expresses $b$.
For $b$ a three-strand braid, it follows that 
$$\Phi(b) = A^{I(b)}I + \Pi(b)$$
\noindent where $I$ is the $ 2 \times 2$ identity matrix and $\Pi(b)$ is a sum of products in the Temperley-Lieb algebra 
involving $U_{1}$ and $U_{2}.$ Since the Temperley-Lieb algebra in this dimension is generated by $I$,$U_{1}$, $U_{2}$,
$U_{1}U_{2}$ and $U_{2}U_{1}$, it follows that the value of the bracket polynomial of the closure of the braid $b$, denoted
$<\overline{b}>,$ can be calculated directly from the trace of this representation, except for the part involving the identity matrix. 
The bracket polynomial evaluation depends upon the loop counts in the states of the closure of the braid, and these loop counts
correspond to the traces of the non-identity Temperley-Lieb elements. Note that the closure of the three-strand diagram for the identity braid 
in $B_{3}$ has bracket polynomial $d^2.$ The
result is the equation 
$$<\overline{b}> = A^{I(b)}d^{2} + tr(\Pi(b))$$
\noindent where $\overline{b}$ denotes the standard braid closure of $b$, and the sharp brackets denote the bracket polynomial. 
Since the trace of the $2 \times 2$ identity matrix is $2$, we see that 
$$<\overline{b}> = tr(\Phi(b)) + A^{I(b)}(d^{2} -2).$$

It follows from this calculation that the question of computing the bracket polynomial for the closure of the three-strand
braid $b$ is mathematically equivalent to the problem of computing the trace of the unitary matrix $\Phi(b).$ 
\vspace{3mm}

\noindent {\bf The Hadamard Test}

In order to (quantum) compute the trace of a unitary matrix $U$, one can use the {\it Hadamard test} to obtain the diagonal matrix
elements $\langle \psi|U|\psi \rangle$ of $U.$ The trace is then the sum of these matrix elements as $|\psi \rangle$ runs over an orthonormal basis for 
the vector space. We first obtain $$\frac{1}{2} + \frac{1}{2}Re\langle \psi|U|\psi \rangle$$ as
an expectation by applying the Hadamard gate $H$
$$H|0 \rangle = \frac{1}{\sqrt{2}}(|0\rangle + |1\rangle)$$
$$H|1 \rangle = \frac{1}{\sqrt{2}}(|0\rangle - |1\rangle)$$
to the first qubit of 
$$C_{U} \circ (H \otimes 1) |0 \rangle |\psi \rangle = \frac{1}{\sqrt{2}}(|0\rangle \otimes|\psi \rangle + |1\rangle \otimes U|\psi\rangle.$$
Here $C_{U}$ denotes controlled $U,$ acting as $U$ when the control bit is $|1 \rangle$ and the identity mapping when the control bit is $|0 \rangle.$ We
measure the expectation for the first qubit $|0 \rangle$ of the resulting state
$$\frac{1}{2}(H|0\rangle \otimes|\psi \rangle + H|1\rangle \otimes U|\psi\rangle)
=\frac{1}{2}((|0\rangle + |1\rangle) \otimes|\psi \rangle + (|0\rangle - |1\rangle) \otimes U|\psi\rangle)$$
$$=\frac{1}{2}(|0\rangle \otimes (|\psi \rangle + U|\psi\rangle) + |1\rangle \otimes(|\psi \rangle - U|\psi\rangle)).$$
This expectation is $$\frac{1}{2}(\langle \psi | + \langle \psi| U^{\dagger})(|\psi \rangle + U|\psi\rangle) = \frac{1}{2} + \frac{1}{2}Re\langle \psi|U|\psi
\rangle.$$
\noindent The imaginary
part is  obtained by applying the same procedure to 
$$\frac{1}{\sqrt{2}}(|0\rangle \otimes|\psi \rangle - i|1\rangle \otimes U|\psi\rangle$$
This is the method used in
\cite{Ah1}, and the reader may wish to contemplate its efficiency in the context of this simple model. Note that the Hadamard test enables this quantum 
computation to estimate the trace of any unitary matrix $U$ by repeated trials that estimate individual matrix entries $\langle \psi|U|\psi\rangle.$
We shall return to quantum algorithms for the Jones polynomial and other knot polynomials in a subsequent paper.
\bigbreak

\section{Quantum Topology, Cobordism Categories, Temperley-Lieb Algebra and Topological Quantum Field Theory}
The purpose of this section is to discuss the general idea behind topological quantum field theory, and to illustrate its application to basic
quantum mechanics and quantum mechanical formalism. It is useful in this regard to have available the concept of {\it category}, and we shall begin the 
section by discussing this far-reaching mathematical concept.
\bigbreak

\noindent {\bf Definition.} A {\it category Cat} consists in two related collections:
\begin{enumerate}
\item $Obj(Cat)$, the {\it objects} of $Cat,$ and 
\item $Morph(Cat)$, the {\it morphisms} of $Cat.$
\end{enumerate}
satisfying the following axioms:
\begin{enumerate}
\item Each morphism $f$ is associated to two objects of $Cat$, the {\it domain} of f and the {\it codomain} of f. Letting $A$ denote the domain of $f$ and
$B$ denote the codomain of $f,$ it is customary to denote the morphism $f$ by the arrow notation
$f:A \longrightarrow B.$
\item Given $f:A \longrightarrow B$ and $g:B \longrightarrow C$ where $A$, $B$ and $C$ are objects of $Cat$, then there exists an associated morphism
$g \circ f : A \longrightarrow C$ called the {\it composition} of $f$ and $g$.
\item To each object $A$ of $Cat$ there is a unique {\it identity morphism} $1_{A}:A \longrightarrow A$ such that $1_{A} \circ f = f$ for any
morphism $f$ with codomain $A$, and $g \circ 1_{A} = g$ for any morphism $g$ with domain $A.$
\item Given three morphisms $f:A \longrightarrow B$, $g:B \longrightarrow C$ and $h:C \longrightarrow D$, then composition is associative.
That is $$(h \circ g) \circ f = h \circ (g \circ f).$$
\end{enumerate}

\noindent If $Cat_{1}$ and $Cat_{2}$ are two categories, then a {\it functor} $F:Cat_{1} \longrightarrow Cat_{2}$ consists in functions 
$F_{O}:Obj(Cat_{1}) \longrightarrow Obj(Cat_{2})$ and $F_{M}:Morph(Cat_{1}) \longrightarrow Morph(Cat_{2})$ such that 
identity morphisms and composition of morphisms are preserved under these mappings. That is (writing just $F$ for $F_{O}$ and $F_{M}$),
\begin{enumerate}
\item $F(1_{A}) = 1_{F(A)}$, 
\item $F(f:A \longrightarrow B) = F(f):F(A) \longrightarrow F(B)$,
\item $F(g \circ f) = F(g) \circ F(f)$.
\end{enumerate}

A functor $F:Cat_{1} \longrightarrow Cat_{2}$ is a structure preserving mapping from one category to another.
It is often convenient to think of the image of the functor $F$ as an {\it interpretation} of the first category in terms of the second.
We shall use this terminology below and sometimes refer to an interpretation without specifying all the details of the functor that describes it.
\bigbreak

The notion of category is a broad mathematical concept, encompassing many fields of mathematics. Thus one has the category of sets where the objects
are sets (collections) and the morphisms are mappings between sets. One has the category of topological spaces where the objects are spaces and the morphisms
are continuous mappings of topological spaces. One has the category of groups where the objects are groups and the morphisms are homomorphisms of groups.
Functors are structure preserving mappings from one category to another. For example, the fundamental group is a functor from the category of topological
spaces with base point, to the category of groups. In all the examples mentioned so far, the morphisms in the category are restrictions of mappings in the 
category of sets, but this is not necessarily the case. For example, any group $G$ can be regarded as a category, $Cat(G)$, with one object $*.$
The morphisms from $*$ to itself are the elements of the group and composition is group multiplication. In this example, the object has no internal structure
and all the complexity of the category is in the morphisms. 
\bigbreak

The Artin braid group $B_{n}$ can be regarded as a category whose single object is an ordered
row of points $[n] = \{1,2,3,...,n \}.$ The morphisms are the braids themselves and composition is the multiplication of the braids. The ordered row of points
is interpreted as the starting and ending row of  points at the bottom and the top of the braid. In the case of the braid category, the morphisms have both
external and internal structure. Each morphism produces a permutation of the ordered row of points (corresponding to the begiinning and ending points of the
individual braid strands), and weaving of the braid is extra structure beyond the object that is its domain and codomain. Finally, for this example, we can
take all the braid groups $B_{n}$ ($n$ a positive integer) under the wing of a single category, $Cat(B)$, whose objects are all ordered rows of points
$[n]$, and whose morphisms are of the form $b:[n] \longrightarrow [n]$ where $b$ is a braid in $B_{n}.$ The reader may wish to have
morphisms between objects with different $n$. We will have this shortly in the Temperley-Lieb category and in the category of tangles.
\bigbreak

The {\it $n$-Cobordism Category}, $Cob[n]$, has as its objects smooth manifolds of dimension $n$, and as its morphisms, smooth manifolds $M^{n+1}$ of
dimension $n+1$ with a partition of the boundary, $\partial M^{n+1}$, into two collections of $n$-manifolds that we denote by $L(M^{n+1})$ and $R(M^{n+1}).$
We regard $M^{n+1}$ as a morphism from  $L(M^{n+1})$ to $R(M^{n+1})$ 
$$M^{n+1}: L(M^{n+1}) \longrightarrow R(M^{n+1}).$$
As we shall see, these cobordism categories are highly significant for 
quantum mechanics, and the simplest one, $Cob[0]$ is directly related to the Dirac notation of bras and kets and to the Temperley-Lieb algebara. We shall
concentrate in this section on these cobordism categories, and their relationships with quantum mechanics.
\bigbreak

\noindent One can choose to consider either oriented or non-oriented manifolds, and within unoriented manifolds there are those that are orientable and 
those that are not orientable. In this section we will implicitly discuss only orientable manifolds, but we shall not specify an orientation. In the 
next section, with the standard definition of topological quantum field theory, the manifolds will be oriented. The definitions of the cobordism
categories for oriented manifolds go over mutatis mutandis.
\bigbreak

Lets begin with $Cob[0]$. Zero dimensional manifolds are just collections of points. The simplest zero dimensional manifold is a single point $p$.
We take $p$ to be an object of this category and also $*$,
where $*$ denotes the empty manifold (i.e. the empty set in the category of manifolds). The object $*$ occurs in $Cob[n]$ for every $n$, since
it is possible that either the left set or the right set of a morphism is empty. A line segment $S$ with boundary points $p$ and $q$ is a morphism from $p$
to $q$.
$$S:p \longrightarrow q$$
See Figure 10. In this figure we have illustrated the morphism from $p$ to $p.$ The simplest convention for this category is to take this morphism to
be the identity. Thus if we look at the subcategory of $Cob[0]$ whose only object is $p$, then the only morphism is the identity morphism. Two points 
occur as the boundary of an interval. The reader will note that $Cob[0]$ and the usual arrow notation for morphisms are very closely related. This is 
a place where notation and mathematical structure share common elements. In general the objects of $Cob[0]$ consist in the empty object $*$
and non-empty rows
of points, symbolized by
$$p \otimes p \otimes \cdots \otimes p \otimes p.$$
Figure 10 also contains a morphism
$$p \otimes p \longrightarrow *$$ and the morphism
$$* \longrightarrow p\otimes p.$$ 
The first represents a cobordism of two points to the empty set (via the bounding curved interval). The second represents a cobordism from the empty set
to two points.

$$ \picill5inby2in(F10)  $$
\begin{center}
{\bf Figure 10 - Elementary Cobordisms}
\end{center}

In Figure 11, we have indicated more morphisms in $Cob[0]$, and we have named the morphisms just discussed as
$$| \Omega \rangle : p \otimes p \longrightarrow *,$$ 
$$\langle \Theta |: * \longrightarrow p\otimes p.$$ 
The point to notice is that the usual conventions for handling Dirac bra-kets are essentially the same as the compostion rules in this
topological category. Thus in Figure 11 we have that
$$\langle \Theta | \circ | \Omega \rangle = \langle \Theta | \Omega \rangle : * \longrightarrow *$$
represents a cobordism from the empty manifold to itself. This cobordism is topologically a circle and, in the Dirac formalism is interpreted as a 
scalar. In order to interpret the notion of scalar we would have to map the cobordism category to the category of vector spaces and linear mappings.
We shall discuss this after describing the similarities with quantum mechanical formalism. Nevertheless, the reader should note that if $V$ is a 
vector space over the complex numbers $C$, then a linear mapping from $C$ to $C$ is determined by the image of $1$, and hence is characterized by the
scalar that is the image of $1$. In this sense a mapping $C \longrightarrow C$ can be regarded as a possible image in vector spaces of the 
abstract structure $\langle \Theta | \Omega \rangle : * \longrightarrow *$. It is therefore assumed that in $Cob[0]$ the composition with the 
morphism $\langle \Theta | \Omega \rangle$ commutes with any other morphism. In that way $\langle \Theta | \Omega \rangle$ behaves like a scalar in 
the cobordism category. In general, an $n+1$ manifold without boundary behaves as a scalar in $Cob[n]$, and if a manifold $M^{n+1}$ can be written
as a union of two submanifolds $L^{n+1}$ and $R^{n+1}$ so that that an $n$-manifold $W^{n}$ is their common boundary:
$$M^{n+1} =  L^{n+1} \cup R^{n+1}$$ with
$$ L^{n+1} \cap R^{n+1} = W^{n}$$ then,
we can write $$\langle M^{n+1} \rangle = \langle L^{n+1} \cup R^{n+1} \rangle = \langle L^{n+1} | R^{n+1} \rangle,$$ and $\langle M^{n+1} \rangle$
will be a scalar (morphism that commutes with all other morphisms) in the category $Cob[n]$.
\bigbreak

$$ \picill5inby3in(F11)  $$
\begin{center}
{\bf Figure 11 - Bras, Kets and Projectors}
\end{center}

$$ \picill5inby3in(F12)  $$
\begin{center}
{\bf Figure 12 - Permutations}
\end{center}

$$ \picill5inby2in(F13)  $$
\begin{center}
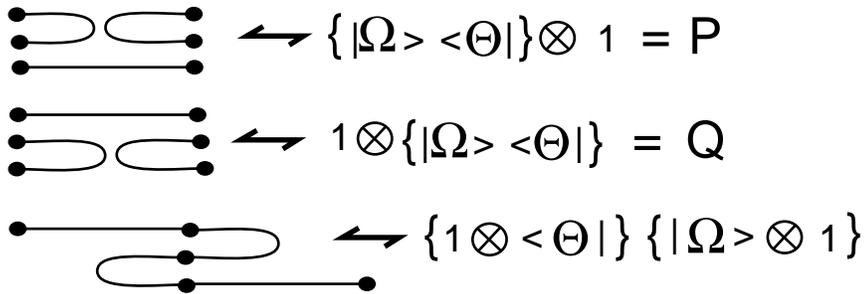
{\bf Figure 13 - Projectors in Tensor Lines and Elementary Topology}
\end{center}

Getting back to the contents of Figure 11, note how the zero dimensional cobordism category has structural parallels to the Dirac ket--bra formalism
$$ U = | \Omega \rangle \langle \Theta |$$
$$ UU = | \Omega \rangle \langle \Theta | \Omega \rangle \langle \Theta |= \langle \Theta | \Omega \rangle | \Omega \rangle \langle \Theta |
= \langle \Theta | \Omega \rangle U.$$ In the cobordism category, the bra--ket and ket--bra formalism is seen as patterns of connection of
the one-manifolds that realize the cobordisms.
\bigbreak

Now view Figure 12. This Figure illustrates a morphism $S$ in $Cob[0]$ that requires two crossed line segments for its planar representation.
Thus $S$ can be regarded as a non-trivial permutation, and $S^2 = I$ where $I$ denotes the identity morphisms for a two-point row.
From this example, it is clear that $Cob[0]$ contains the structure of all the symmetric groups and more. In fact, if we take the subcateogry of 
$Cob[0]$ consisting of all morphisms from $[n]$ to $[n]$ for a fixed positive integer $n,$ then this gives the well-known {\it Brauer algebra}  (see
\cite{Benkart}) extending the symmetric group by allowing any connections among the points in the two rows. In this sense, one could call $Cob[0]$ the {\it
Brauer category}. We shall return to this point of view later.
\bigbreak

In this section, we shall be concentrating
on the part of $Cob[0]$ that does not involve permutations. This part can be characterized by those morphisms that can be represented by 
planar diagrams without crosssings between any of the line segments (the one-manifolds). We shall call this crossingless subcategory of $Cob[0]$ the {\em
Temperley-Lieb Category} and denote it by $CatTL.$ In $CatTL$ we have the subcategory $TL[n]$ whose only objects are the row of $n$ points and the empty
object $*$, and whose morphisms can all be represented by configurations that embed in the plane as in the morphisms $P$ and $Q$ in Figure 13. Note that with
the empty object $*$, the morphism whose diagram is a single loop appears in $TL[n]$ and is taken to commute with all other morphisms.
\bigbreak

The {\em Temperley-Lieb Algebra}, $AlgTL[n]$ is generated by the morphisms in $TL[n]$ that go from $[n]$ to itself. 
Up to multiplication by the loop, the product (composition) of two such morphisms is another flat morphism from $[n]$ to itself.
 For algebraic purposes the loop $*
\longrightarrow *$ is taken to be a scalar  algebraic variable $\delta$ that commutes with all elements in the algebra. Thus the equation 
$$ UU = \langle \Theta | \Omega \rangle U.$$
becomes
$$UU = \delta U$$
in the algebra. In the algebra we are allowed to add morphisms formally and this addition is taken to be commutative. Initially the algebra is taken with 
coefficients in the integers, but a different commutative ring of coefficients can be chosen and the value of the loop may be taken in this ring. For example,
for quantum mechanical applications it is natural to work over the complex numbers. The multiplicative structure of $AlgTL[n]$ can be described by 
generators and relations as follows: Let $I_{n}$ denote the identity morphism from $[n]$ to $[n].$ Let $U_{i}$ denote the morphism from $[n]$ to $[n]$
that connects $k$ with $k$ for $k<i$ and $k>i+1$ from one row to the other, and connects $i$ to $i+1$ in each row. Then the algebra
$AlgTL[n]$ is generated by $\{ I_{n}, U_{1},U_{2},\cdots ,U_{n-1} \}$ with relations
$$U_{i}^{2} = \delta U_{i}$$
$$U_{i}U_{i+1}U_{i} = U_{i}$$
$$U_{i}U_{j} = U_{j}U_{i} \,: \,\, |i-j|>1.$$
These relations are illustrated for three strands in Figure 13. We leave the commuting relation for the reader to draw in the case where $n$ is 
four or greater. For a proof that these are indeed all the relations, see \cite{KaufDiag}.
\bigbreak

Figures 13 and 14 indicate how the zero dimensional cobordism category contains structure that goes well beyond the usual Dirac formalism.
By tensoring the ket--bra on one side or another by identity morphisms, we obtain the beginnings of the Temperley-Lieb algebra and the Temperley-Lieb
category. Thus Figure 14 illustrates the morphisms $P$ and $Q$ obtained by such tensoring, and the relation $PQP = P$ which is the same as
$U_{1}U_{2}U_{1} = U_{1}$
\bigbreak

Note the composition at the 
bottom of the Figure 14. Here we see a composition of the identity tensored with a ket, followed by a bra tensored with the identity.
The diagrammatic for this
association involves ``straightening" the curved structure of the morphism to a straight line. 
In Figure 15 we have elaborated this
situation even further, pointing out that in this category each of the morphisms $\langle \Theta |$ and $| \Omega \rangle$ can be seen, by 
straightening, as mappings from 
the generating object to itself. We have denoted these corresponding morphisms by $\Theta$ and $\Omega$ respectively.
In this way there is a correspondence between
morphisms $p \otimes p \longrightarrow *$ and morphims $p \longrightarrow p.$ 
\bigbreak

In Figure 15 we have illustrated the generalization of the straightening
procedure of Figure 14. In Figure 14 the straightening occurs because the connection structure in the morphism of $Cob[0]$ does not depend on 
the wandering of curves in diagrams for the morphisms in that category. Nevertheless, one can envisage a more complex interpretation of the 
morphisms where each one-manifold (line segment) has a label, and a multiplicity of morphisms can correspond to a single line segment.
This is exactly what we expect in interpretations. For example, we can interpret the line segment $[1] \longrightarrow [1]$ as a mapping from 
a vector space $V$ to itself. Then $[1] \longrightarrow [1]$ is the diagrammatic abstraction for $ V \longrightarrow V,$ and there are many
instances of linear mappings from $V$ to $V$. 
\bigbreak

At the vector space level there is a duality between mappings 
$V \otimes V \longrightarrow C$  and linear maps $V \longrightarrow V.$
Specifically, let 
$$\{ | 0 \rangle ,\cdots, | m \rangle \}$$
be a basis for $V.$ Then $\Theta: V \longrightarrow V$ is determined by
$$\Theta |i \rangle = \Theta_{ij} \, |j \rangle$$ (where we have used the Einstein summation convention on the repeated index $j$) 
corresponds to the bra 
$$\langle \Theta |: V \otimes V \longrightarrow C$$
defined by 
$$\langle \Theta |ij \rangle = \Theta_{ij}.$$
Given $\langle \Theta | :V \otimes V \longrightarrow C,$
we associate $\Theta: V \longrightarrow V$ in this way. 
\bigbreak

Comparing with the diagrammatic for the category $Cob[0]$, we say that $\Theta: V \longrightarrow V$ is obtained by {\it straightening}
the mapping $$\langle \Theta | :V \otimes V \longrightarrow C.$$ Note that in this interpretation, the bras and kets are defined relative to the 
tensor product of $V$ with itself and $[2]$ is interpreted as $V \otimes V.$ If we interpret $[2]$ as a single vector space $W,$ then 
the usual formalisms of bras and kets still pass over from the cobordism category.
\bigbreak

$$ \picill5inby3in(F14)  $$
\begin{center}
{\bf Figure 14 - The Basic Temperley-Lieb Relation}
\end{center}

$$ \picill3inby4.5in(F15)  $$
\begin{center}
{\bf Figure 15 - The Key to Teleportation}
\end{center}

Figure 15 illustrates the staightening of $| \Theta \rangle$ and $\langle \Omega |,$ and the straightening of a composition of these
applied to $| \psi \rangle,$ resulting in $| \phi \rangle.$ In the left-hand part of the bottom of Figure 15 we illustrate the preparation
of the tensor product $| \Theta \rangle \otimes | \psi \rangle$ followed by a successful measurement by $\langle \Omega |$ in the second two
tensor factors. The resulting single qubit state, as seen by straightening, is  $| \phi \rangle = \Theta \circ \Omega |\psi \rangle.$ 
\bigbreak

From this, we see that it is possible to reversibly, indeed unitarily, transform a state $| \psi \rangle$ via a combination of preparation and measurement
just so long as the straightenings of the preparation and measurement ($\Theta$ and $\Omega$) are each invertible (unitary). This is the 
key to teleportation \cite{Teleport,C1,C2}. In the standard teleportation procedure one chooses the preparation $\Theta$ to be (up to normalization) the  $2$
dimensional identity matrix so that
$| \theta \rangle = |00\rangle + |11\rangle.$ If the successful measurement $\Omega$ is also the identity, then the transmitted state $| \phi \rangle$
will be equal to $| \psi \rangle.$ In general we will have $| \phi \rangle =  \Omega |\psi \rangle.$ One can then choose a basis of measurements
$|\Omega \rangle,$ each corresponding to a unitary transformation $\Omega$ so that the recipient of the transmission can rotate the result by the 
inverse of $\Omega$ to reconsitute $|\psi \rangle$ if he is given the requisite information. This is the basic design of the teleportation procedure.
\bigbreak  

There is much more to say about the category $Cob[0]$ and its relationship with quantum mechanics. We will stop here, and invite the reader to explore
further. Later in this paper, we shall use these ideas in formulating our representations of the braid group. For now, we point out how things look
as we move upward to $Cob[n]$ for $n > 0.$ In Figure 16 we show typical cobordisms (morphisms) in $Cob[1]$ from two circles to one circle and from 
one circle to two circles. These are often called ``pairs of pants". Their composition is a surface of genus one seen as a morphism from two circles
to two circles. The bottom of the figure indicates a ket-bra in this dimension in the form of a mapping from one circle to one circle as a composition of
a cobordism of a circle to the empty set and a cobordism from the empty set to a circle (circles bounding disks). As we go to higher dimensions the 
structure of cobordisms becomes more interesting and more complicated. It is remarkable that there is so much structure in the lowest dimensions of 
these categories.

$$ \picill2inby3.5in(F16)  $$
\begin{center}
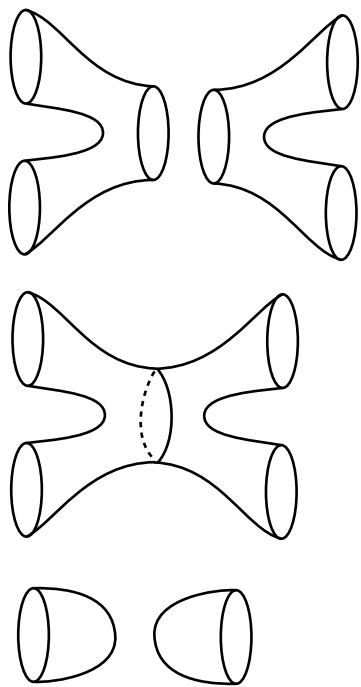
{\bf Figure 16 - Corbordisms of $1$-Manifolds are Surfaces}
\end{center}

\section{Braiding and Topological Quantum Field Theory}
The purpose of this section is to discuss in a very general way how braiding is related to topological quantum field theory. 
In the section to follow, we will use the Temperley-Lieb recoupling theory to produce specfic unitary representations of the Artin
braid group.
\bigbreak 

The ideas in the subject
of topological quantum field theory (TQFT) are well expressed in the book \cite{Atiyah} by Michael Atiyah and the paper \cite{Witten} by Edward Witten.
Here is Atiyah's definition:
\bigbreak

\noindent {\bf Definition.} A TQFT in dimension $d$ is a functor $Z(\Sigma)$ from the cobordism category $Cob[d]$ to the category  $Vect$ of
vector spaces and linear mappings which assigns
\begin{enumerate}
\item a finite dimensional vector space $Z(\Sigma)$ to each compact, oriented $d$-dimensional manifold $\Sigma,$ 
\item a vector $Z(Y) \in Z(\Sigma)$ for each compact, oriented $(d + 1)$-dimensional manifold $Y$ with boundary $\Sigma.$
\item a linear mapping $Z(Y):Z(\Sigma_{1}) \longrightarrow Z(\Sigma_{2})$ when $Y$ is a $(d + 1)$-manifold that is a cobordism 
between $\Sigma_{1}$ and $\Sigma_{2}$ (whence the boundary of $Y$ is the union of $\Sigma_{1}$ and $-\Sigma_{2}.$
\end{enumerate}

\noindent The functor satisfies the following axioms.

\begin{enumerate}
\item $Z(\Sigma^{\dagger}) = Z(\Sigma)^{\dagger}$ where $\Sigma^{\dagger}$ denotes the manifold $\Sigma$ with the opposite orientation and 
$Z(\Sigma)^{\dagger}$ is the dual vector space.
\item $Z(\Sigma_{1} \cup \Sigma_{2}) = Z(\Sigma_{1}) \otimes Z(\Sigma_{2})$ where $\cup$ denotes disjoint union.
\item If $Y_{1}$ is a cobordism from $\Sigma_{1}$ to $\Sigma_{2},$  $Y_{2}$ is a cobordism from $\Sigma_{2}$ to $\Sigma_{3}$ and 
$Y$ is the composite cobordism $Y = Y_{1} \cup_{\Sigma_{2}} Y_{2},$ then 
$$Z(Y) = Z(Y_{2}) \circ Z(Y_{1}): Z(\Sigma_{1}) \longrightarrow Z(\Sigma_{2})$$ is the composite of the corresponding linear mappings.
\item $Z(\phi) = C$ ($C$ denotes the complex numbers) for the empty manifold $\phi.$
\item With $\Sigma \times I$ (where $I$ denotes the unit interval) denoting the identity cobordism from $\Sigma$ to $\Sigma,$ 
$Z(\Sigma \times I)$ is the identity mapping on $Z(\Sigma).$
\end{enumerate}

Note that, in this view a TQFT is basically a functor from the cobordism categories defined in the last section to Vector Spaces
over the complex numbers. We have already seen that in the lowest dimensional case of cobordisms of zero-dimensional manifolds, this gives
rise to a rich structure related to quatum mechanics and quantum information theory. The remarkable fact is that the case of three-dimensions
is also related to quantum theory, and to the lower-dimensional versions of the TQFT. This gives a significant way to think about three-manifold
invariants in terms of lower dimensional patterns of interaction. Here follows a brief description.
\bigbreak

Regard the three-manifold as a union of two handlebodies with boundary an orientable  surface $S_{g}$ of genus $g.$ The surface is divided up into trinions as
illustrated in Figure 17. A {\it trinion} is a surface with boundary that is topologically equivalent to a sphere with three punctures. The trinion
constitutes, in itself a cobordism in $Cob[1]$ from two circles to a single circle, or from a single circle to two circles, or from three circles to the 
empty set. The {\it pattern} of a trinion is a trivalent graphical vertex, as illustrated in Figure 17. In that figure we show the trivalent vertex
graphical pattern drawn on the surface of the trinion, forming a graphical pattern for this combordism. It should be clear from this figure that any
cobordism in $Cob[1]$ can be diagrammed by a trivalent graph, so that the category of trivalent graphs (as morphisms from ordered sets of points to ordered
sets of points) has an image in the category of cobordisms of  compact one-dimensional manifolds. Given a surface $S$ (possibly with boundary) and a
decomposition of that surface into triions, we associate to it a trivalent graph
$G(S,t)$ where $t$ denotes the particular trinion decomposition.
\bigbreak

In this correspondence, distinct graphs can correspond to topologically identical cobordisms of circles, as
illustrated in Figure 19. It turns out that the graphical structure is important, and that it is extraordinarily useful to articulate transformations
between the graphs that correspond to the homeomorphisms of the corresponding surfaces. The beginning of this structure is indicated in the bottom part of
Figure 19. 
\bigbreak

In Figure 20 we illustrate another feature of the relationship betweem surfaces and graphs. At the top of the figure we indicate a 
homeomorphism between a twisted trinion and a standard trinion. The homeomorphism leaves the ends of the trinion (denoted $A$,$B$ and $C$) fixed while undoing
the internal twist. This can be accomplished as an ambient isotopy of the embeddings in three dimensional space that are indicated by this figure.
Below this isotopy we indicate the corresponding graphs. In the graph category there will have to be a transformation between a braided and an unbraided
trivalent vertex that corresponds to this homeomorphism.
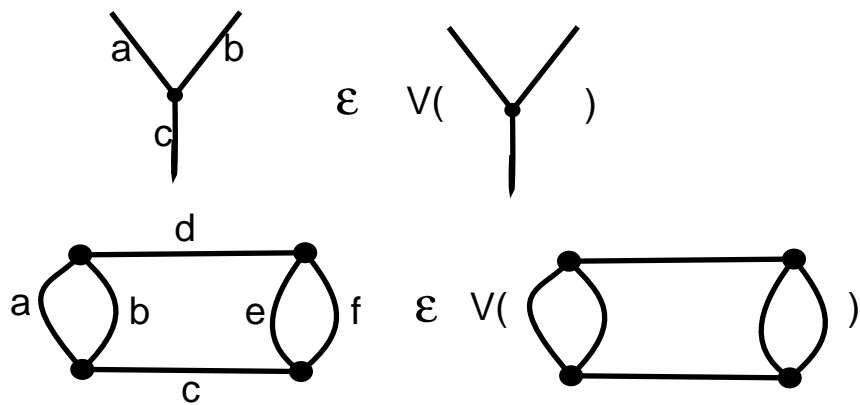
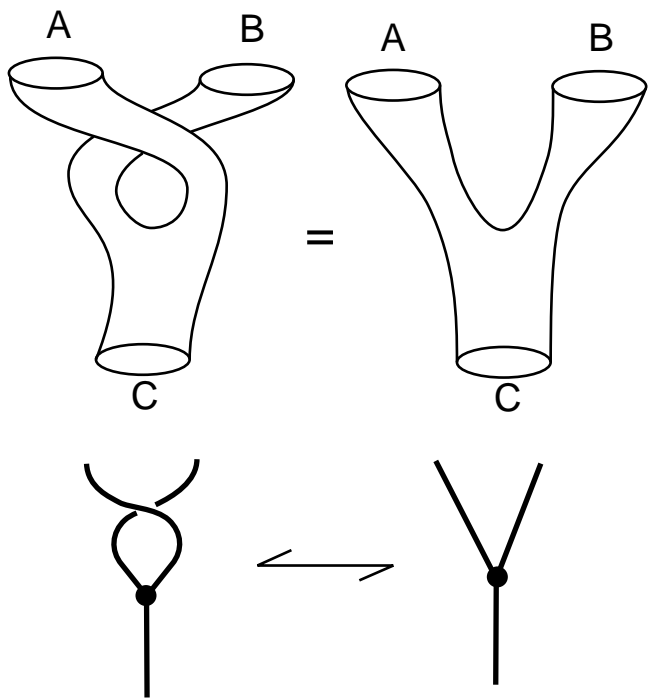
\bigbreak

$$ \picill3inby4in(F17)  $$
\begin{center}
{\bf Figure 17 - Decomposition of a Surface into Trinions}
\end{center}

$$ \picill3inby2.5in(F18)  $$
\begin{center}
{\bf Figure 18 - Trivalent Vectors}
\end{center}

$$ \picill2.5inby3.5in(F19)  $$
\begin{center}
{\bf Figure 19 - Trinion Associativity}
\end{center}

$$ \picill2.5inby3.5in(F20)  $$
\begin{center}
{\bf Figure 20 - Tube Twist}
\end{center}

From the point of view that we shall take in this paper, the key to the mathematical structure of three-dimensional TQFT lies in the trivalent graphs,
including the  braiding of grapical arcs. We can think of these braided graphs as representing idealized Feynman diagrams,
with the trivalent vertex as the basic particle interaction vertex, and the braiding of lines representing an interaction resulting from an exchange of 
particles. In this view one thinks of the particles as moving in a two-dimensional medium, and the diagrams of braiding and trivalent vertex interactions
as indications of the temporal events in the system, with time indicated in the direction of the morphisms in the category. Adding such graphs to the category
of knots and links is an extension of the {\it tangle category} where one has already extended braids to allow any embedding of strands and circles that 
start in $n$ ordered points and end in $m$ ordered points. The tangle category includes the braid category and the Temperley-Lieb category. These are 
both included in the category of braided trivalent graphs. 
\bigbreak

Thinking of the basic trivalent vertex as the form of a particle interaction there will be a set of particle states that can label each arc incident to 
the vertex. In Figure 18 we illustrate the labeling of the trivalent graphs by such particle states. In the next two sections we will see specific rules
for labeling such states. Here it suffices to note that there will be some restrictions on these labels, so that a trivalent vertex has a set of possible
labelings. Similarly, any trivalent graph will have a set of admissible labelings. These are the possible particle processes that this graph can support.
We take the set of admissible labelings of a given graph $G$ as a basis for a vector space $V(G)$ over the complex numbers. This vector space is the space
of {\it processes} associated with the graph $G.$ Given a surface $S$ and a decomposition $t$ of the surface into trinions, we have the associated 
graph $G(S,t)$ and hence a vector space of processes $V(G(S,t))$. It is desirable to have this vector space independent of the particular decomposition 
into trinions. If this can be accomplished, then the set of vector spaces and linear mappings associated to the surfaces can consitute a functor from the
category of cobordisms of  one-manifolds to vector spaces, and hence gives rise to a  one-dimensional topological quantum field theory. To this end we need
some properties of the particle interactions that will be described below. 
\bigbreak

A {\it spin network} is, by definition a lableled trivalent graph in a category of graphs that satisfy the properties outlined in the previous
paragraph. We shall detail the requirements below. 
\bigbreak 

The simplest  case of this idea is C. N. Yang's original interpretation of the Yang-Baxter Equation \cite{Yang}.  Yang articulated a
quantum field theory in one dimension of space and one dimension of time in which the $R$-matrix giving the
scattering ampitudes for an interaction of two particles whose (let us say) spins corresponded  to the matrix indices so that
$R^{cd}_{ab}$ is the amplitude for particles of spin $a$ and spin $b$ to interact and produce particles of spin $c$ and $d.$ Since these interactions are
between particles in a line, one takes the convention that the particle with spin
$a$ is to the left of the particle with spin $b,$ and the particle with spin $c$ is to the left of the particle with spin $d.$
If one follows the concatenation of such interactions, then there is an underlying permutation that is obtained
by following strands from the bottom to the top of the diagram (thinking of time as moving up the page). Yang designed the 
Yang-Baxter equation for $R$ so that {\em the amplitudes for a composite process depend only on the underlying permutation corresponding to the
process and not on the individual sequences of interactions.} 
\bigbreak

In taking over the Yang-Baxter equation for topological purposes, we can use the same interpretation, but think of the diagrams with 
their under- and over-crossings as modeling events in a spacetime with two dimensions of space and one dimension of time. The extra
spatial dimension is taken in displacing the woven strands perpendicular to the page, and allows us to use braiding operators $R$ and
$R^{-1}$ as  scattering matrices. Taking this picture to heart, one can add other particle properties to the idealized theory. In
particular one can  add fusion and creation vertices where in fusion two particles interact to become a single particle and in creation
one particle  changes (decays) into two particles. These are the trivalent vertices discussed above. Matrix elements corresponding to trivalent vertices can
represent these interactions. See Figure 21.
\bigbreak

$$ \picill3inby1in(F21)  $$

\begin{center}
{\bf Figure 21 -Creation and Fusion}
\end{center}

Once one introduces trivalent vertices for fusion and creation, there is the question how these interactions will behave in respect to 
the braiding operators. There will be a matrix expression for the compositions of braiding and fusion or creation as indicated in Figure
22. Here we  will restrict ourselves to showing the diagrammatics with the intent of giving the reader a flavor of these
structures. It is natural to assume that braiding intertwines with creation as shown in Figure 24 (similarly with fusion). This
intertwining identity is clearly the sort of thing that a topologist will love, since it indicates that the diagrams can be interpreted
as embeddings of graphs in three-dimensional space, and it fits with our interpretation of the vertices in terms of trinions. Figure 22 illustrates the
Yang-Baxter equation.  The intertwining identity is an assumption like the
Yang-Baxter equation itself, that simplifies the mathematical structure of the model.
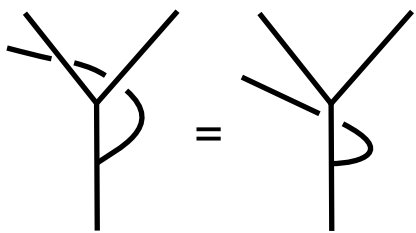
\bigbreak

$$ \picill3inby1.5in(F22)  $$

\begin{center}
{\bf Figure 22 - YangBaxterEquation}
\end{center}

$$ \picill3inby1.5in(F23)  $$

\begin{center}
{\bf Figure 23 - Braiding}
\end{center}

$$ \picill3inby1.5in(F24)  $$

\begin{center}
{\bf Figure 24 - Intertwining}
\end{center}

It is to be expected that there will be an operator that expresses the recoupling of vertex interactions as shown in Figure 25 and labeled
by $Q.$  This corresponds to the associativity at the level of trinion combinations shown in Figure 19. The actual formalism of such an operator will
parallel the mathematics of recoupling for angular momentum. See for example 
\cite{KL}. If one just considers the abstract structure of recoupling then one sees that for trees with four branches (each with a single
root) there is a cycle of length five as shown in Figure 26. One can start with any pattern of three vertex interactions and 
go through a sequence of five recouplings that bring one back to the same tree from which one started. {\em It is a natural simplifying 
axiom to assume that this composition is the identity mapping.} This axiom is called the {\em pentagon identity}. 
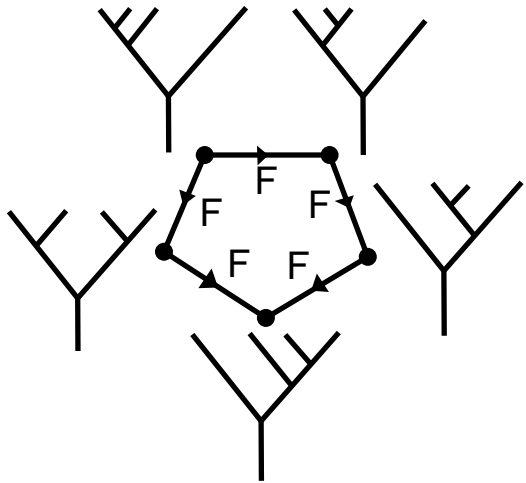
\bigbreak

$$ \picill3inby1.5in(F25)  $$

\begin{center}
{\bf Figure 25 - Recoupling}
\end{center}

$$ \picill3inby3in(F26)  $$

\begin{center}
{\bf Figure 26 - Pentagon Identity}
\end{center}

Finally there is a hexagonal cycle of interactions between braiding, recoupling and the intertwining identity as shown in Figure 27.
One says that the interactions satisfy the {\em hexagon identity} if this composition is the identity.
\bigbreak

$$ \picill3inby4in(F27)  $$

\begin{center}
{\bf Figure 27 - Hexagon Identity}
\end{center}

A {\em graphical three-dimensional topological quantum field theory} is an algebra of interactions that satisfies the Yang-Baxter equation, the
intertwining identity, the pentagon identity  and the hexagon identity. There is not room in this summary to detail the way
that these properties fit into the topology of knots and three-dimensional manifolds, but a sketch is in order. For the case of topological 
quantum field theory related to the group $SU(2)$ there is a construction based entirely on the combinatorial topology of the bracket polynomial
(See Sections 7,9 and 10 of this article.). See \cite{KP,KL} for more information on this approach.
\bigbreak

Now return to Figure 17 where we
illustrate trinions, shown in relation to  a trivalent vertex, and a surface of genus three that is decomposed into four trinions. It
turns out that the vector space
$V(S_g) = V(G(S_{g},t))$ to a surface with a trinion decomposition as $t$ described above, and defined in terms of the graphical topological quantum field
theory, does not depend upon the choice of trinion decomposition. This independence is guaranteed by
the braiding, hexagon and pentagon identities. One can then associate a well-defined vector $|M \rangle$ in $V(S_{g})$ whenenver $M$
is a three manifold whose  boundary is $S_{g}.$  Furthermore, if a closed three-manifold $M^{3}$ is decomposed along a surface $S_{g}$ into 
the union of $M_{-}$ and $M_{+}$
where these parts are otherwise disjoint three-manifolds with boundary $S_{g},$ then the inner product $I(M) = \langle M_{-} | M_{+} \rangle$ is, up to
normalization, an invariant of the three-manifold $M_{3}.$ With the definition of graphical topological quantum field theory given above, knots and links can
be incorporated as well, so that one obtains a source of invariants $I(M^{3},K)$ of knots and links in orientable three-manifolds. Here we see the uses of
the relationships that occur in the higher dimensional cobordism categories, as descirbed in the previous section.
\bigbreak 

\noindent The invariant $I(M^{3},K)$ can be formally compared with the Witten \cite{Witten} integral  $$Z(M^{3},K) = \int DAe^{(ik/4\pi)S(M,A)} W_{K}(A).$$ It
can be shown that up to limits of the heuristics, $Z(M,K)$ and $I(M^{3},K)$ are essentially equivalent for appropriate choice of gauge group and
corresponding spin networks.
\bigbreak

By these graphical reformulations, a three-dimensional $TQFT$ is, at base, a highly simplified theory of point particle interactions in $2+1$
dimensional spacetime. It can be used to articulate invariants of knots and links and invariants of three manifolds. The reader
interested in the
$SU(2)$ case of this structure and its implications for invariants of knots and three manifolds can consult \cite{KL,KP,Kohno,Crane,MS}. One expects that
physical situations involving
$2+1$ spacetime will be approximated by such an idealized theory.  There are also applications to $3 + 1$ quantum gravity \cite{ASR,AL,KaufLiko}.
Aspects of the quantum Hall effect may be related to topological quantum field theory
\cite{Wilczek}. One can study a physics in two dimensional space where the braiding of  particles or collective excitations leads to non-trival
representations of the Artin braid group. Such particles are called {\it Anyons}.  Such $TQFT$ models would describe applicable physics. One can
think about applications of anyons to quantum computing along the lines of the topoological models described here.  
\bigbreak

$$ \picill3inby2.5in(F28)  $$
\begin{center}
{\bf Figure 28 - A More Complex Braiding Operator}
\end{center}
\bigbreak

A key point in the application of $TQFT$ to quantum information theory is contained in the 
structure illustrated in Figure 28. There we show a more complex braiding operator, based on the composition of recoupling with the
elementary braiding at a vertex. (This structure is implicit in the Hexagon identity of Figure 27.) The new braiding operator is a 
source of unitary representations of braid group in situations (which exist mathematically) where the recoupling transformations are themselves 
unitary. This kind of pattern is utilized in the work of Freedman and collaborators \cite{F,FR98,FLZ,Freedman5,Freedman6}
and in the case of classical angular momentum formalism has been dubbed a ``spin-network quantum simlator" by Rasetti and collaborators
\cite{MR,MR2}. In the next section we show how certain natural deformations \cite{KL} of Penrose spin networks \cite{Penrose} can be used 
to produce these unitary representations of the Artin braid group  and the corresponding models for anyonic topological quantum computation. 
\bigbreak

\section {Spin Networks and Temperley-Lieb Recoupling Theory}
In this section we discuss a combinatorial construction for spin networks that generalizes the original construction of Roger Penrose.
The result of this generalization is a structure that satisfies all the properties of a graphical $TQFT$ as described in the previous section, and 
specializes to classical angular momentum recoupling theory in the limit of its basic variable. The construction is based on the properties of 
the bracket polynomial (as already described in Section 4). A complete description of this theory can be found in the book ``Temperley-Lieb
Recoupling Theory and Invariants of Three-Manifolds" by Kauffman and Lins \cite{KL}.  
\bigbreak

The ``$q$-deformed" spin networks that we construct here are based on the bracket polynomial relation. View Figure 29 and Figure 30.
\bigbreak

$$ \picill5inby4in(F29)  $$

\begin{center}
{\bf Figure 29 - Basic Projectors }
\end{center}
\bigbreak

$$ \picill5inby3in(F30)  $$

\begin{center}
{\bf Figure 30 - Two Strand Projector}
\end{center}
\bigbreak

$$ \picill5inby3in(F31)  $$
\begin{center}
{\bf Figure 31 -Vertex}
\end{center}
\bigbreak

In Figure 29 we indicate how the basic projector (symmetrizer, Jones-Wenzl projector) $$ \picill.25inby.25in(symm) $$
\bigbreak

\noindent is constructed on the basis of the
bracket polynomial expansion. In this technology a symmetrizer is a sum of tangles on $n$ strands (for a chosen integer $n$). The tangles are made by
summing over braid lifts of  permutations in the symmetric group on $n$ letters, as indicated in Figure 29. Each elementary braid is then expanded by the
bracket polynomial  relation as indicated in Figure 29 so that the resulting sum  consists of flat tangles without any crossings (these can be viewed as
elements in the Temperley-Lieb algebra). The projectors have the property that the concatenation of a projector with itself is just that projector, and
if you tie two lines on the top or the bottom of a projector together, then the evaluation is zero. This general definition of projectors is very useful for 
this theory. The two-strand projector is shown in Figure 30. Here the formula for that projector 
is particularly simple. It is the sum of two parallel arcs and two turn-around arcs (with coefficient $-1/d,$  with $d = -A^{2} - A^{-2}$ is the loop
value for the bracket polynomial. Figure 30 also shows the recursion formula for the general projector. This recursion formula is due to Jones and Wenzl and
the projector in this form, developed as a sum in the Temperley--Lieb algebra (see Section 5 of this paper), is usually known as the {\em Jones--Wenzl
projector}.
\bigbreak

The projectors are combinatorial analogs of irreducible representations of a group (the original spin nets were based
on $SU(2)$ and these deformed nets are based on the corresponding quantum group to SU(2)). As such the reader can think of them as ``particles". The
interactions of these particles are governed by how they can be tied together into three-vertices. See Figure 31.
In Figure 31 we show how to tie three projectors, of $a,b,c$ strands respectively, together to form a three-vertex. In order to accomplish this 
interaction, we must share lines between them as shown in that Figure so that there are non-negative integers $i,j,k$ so that
$a = i + j, b = j + k, c = i + k.$ This is equivalent to the condition that $a + b + c$ is even and that the sum of any two of $a,b,c$ is 
greater than or equal to the third. For example $a + b \ge c.$ One can think of the vertex as a possible particle interaction where
$[a]$ and $[b]$ interact to produce $[c].$ That is, any two of the legs of the vertex can be regarded as interacting to produce the third leg.
\bigbreak

There is a basic orthogonality of three vertices as shown in Figure 32. Here if we tie two three-vertices together
so that they form a ``bubble" in the middle, then the resulting network with labels $a$ and $b$ on its free ends
is a multiple of an $a$-line (meaning a line with an $a$-projector on it) or zero (if $a$ is not equal to $b$).
The multiple is compatible with the results of closing the diagram in the equation of Figure 32 so the two free
ends are identified with one another. On closure, as shown in the Figure, the left hand side of the equation becomes
a Theta graph and the right hand side becomes a multiple of a ``delta" where $\Delta_{a}$ denotes the bracket 
polynomial evaluation of the $a$-strand loop with a projector on it. The $\Theta(a,b,c)$ denotes the bracket 
evaluation of a theta graph made from three trivalent vertices and labeled with $a, b, c$ on its edges.
\bigbreak

There is a recoupling formula in this theory in the form shown in Figure 33.
Here there are ``$6$-j symbols", recoupling coefficients that can be expressed, as shown in 
Figure 35, in terms of tetrahedral graph evaluations and theta graph evaluations. The tetrahedral graph is shown in 
Figure 34. One derives the formulas for 
these coefficients directly from the orthogonality relations for the trivalent vertices by 
closing the left hand side of the recoupling formula and using orthogonality to evaluate the right hand side.
This is illustrated in Figure 35.

$$ \picill5inby5in(F32)  $$

\begin{center}
{\bf Figure 32 - Orthogonality of Trivalent Vertices}
\end{center}
\bigbreak

 $$ \picill5inby1.2in(F33)  $$

\begin{center}
{\bf Figure 33 - Recoupling Formula}
\end{center}
\bigbreak

$$ \picill5inby1in(F34)  $$

\begin{center}
{\bf Figure 34 - Tetrahedron Network}
\end{center}
\bigbreak

$$ \picill3inby3.5in(F35)  $$

\begin{center}
{\bf Figure 35 - Tetrahedron Formula for Recoupling Coefficients}
\end{center}
\bigbreak

Finally, there is the braiding relation, as illustrated in Figure 36.

$$ \picill3inby3in(F36)  $$

\begin{center}
{\bf Figure 36 - Local Braiding Formula}
\end{center}
\bigbreak

With the braiding relation in place, this $q$-deformed spin network theory satisfies the pentagon, hexagon and braiding naturality identities
needed for a topological quantum field theory. All these identities follow naturally from the basic underlying topological construction of the 
bracket polynomial. One can apply the theory to many different situations.

\subsection{Evaluations}
In this section we discuss the structure of the evaluations for $\Delta_{n}$ and the theta and tetrahedral networks. We refer to 
\cite{KL} for the details behind these formulas. Recall that $\Delta_{n}$ is the bracket evaluation of the closure of the $n$-strand
projector, as illustrated in Figure 32. For the bracket variable $A,$ one finds that 
$$\Delta_{n} = (-1)^{n}\frac{A^{2n+2} - A^{-2n-2}}{A^{2} - A^{-2}}.$$
One sometimes writes the {\it quantum integer}
$$[n] = (-1)^{n-1}\Delta_{n-1} = \frac{A^{2n} - A^{-2n}}{A^{2} - A^{-2}}.$$
If $$A=e^{i\pi/2r}$$ where $r$ is a positive integer, then 
$$\Delta_{n} = (-1)^{n}\frac{sin((n+1)\pi/r)}{sin(\pi/r)}.$$
Here the corresponding quantum integer is
$$[n] = \frac{sin(n\pi/r)}{sin(\pi/r)}.$$
Note that $[n+1]$ is a positive real number for $n=0,1,2,...r-2$ and that $[r-1]=0.$
\bigbreak

The evaluation of the theta net is expressed in terms of quantum integers by the formula
$$\Theta(a,b,c) = (-1)^{m + n + p}\frac{[m+n+p+1]![n]![m]![p]!}{[m+n]![n+p]![p+m]!}$$
where $$a=m+p, b=m+n, c=n+p.$$ Note that $$(a+b+c)/2 = m + n + p.$$
\bigbreak

When $A=e^{i\pi/2r},$ the recoupling theory becomes finite with the restriction that only three-vertices
(labeled with $a,b,c$) are {\it admissible} when $a + b +c \le 2r-4.$ All the summations in the 
formulas for recoupling are restricted to admissible triples of this form.
\bigbreak

\subsection{Symmetry and Unitarity}
The formula for the recoupling coefficients given in Figure 35 has less symmetry than is actually inherent in the structure of the situation.
By multiplying all the vertices by an appropriate factor, we can reconfigure the formulas in this theory so that the revised recoupling transformation is
orthogonal, in the sense that its transpose is equal to its inverse. This is a very useful fact. It means that when the resulting matrices are real, then
the recoupling transformations are unitary. We shall see particular applications of this viewpoint later in the paper.
\bigbreak

Figure 37 illustrates this modification of the three-vertex. Let $Vert[a,b,c]$ denote the original $3$-vertex of the Temperley-Lieb recoupling theory.
Let $ModVert[a,b,c]$ denote the modified vertex. Then we have the formula
$$ModVert[a,b,c] = \frac{\sqrt{\sqrt{\Delta_{a} \Delta_{b} \Delta_{c}}}}{ \sqrt{\Theta(a,b,c)}}\,\, Vert[a,b,c].$$

\noindent {\bf Lemma.}  {\it For the bracket evaluation at the root of unity $A = e^{i\pi/2r}$ the factor
$$f(a,b,c) = \frac{\sqrt{\sqrt{\Delta_{a} \Delta_{b} \Delta_{c}}}}{ \sqrt{\Theta(a,b,c)}}$$
is real, and can be taken to be a positive real number for $(a,b,c)$ admissible (i.e. $a + b + c \le 2r -4$).}
\bigbreak

\noindent {\bf Proof.} By the results from the previous subsection, 
$$\Theta(a,b,c) = (-1)^{(a+b+c)/2}\hat{\Theta}(a,b,c)$$ where $\hat{\Theta}(a,b,c)$ is positive real, and 
$$\Delta_{a} \Delta_{b} \Delta_{c} = (-1)^{(a+b+c)} [a+1][b+1][c+1]$$ where the quantum integers in this formula can be taken to be
positive real. It follows from this that
$$f(a,b,c) = \sqrt{\frac{\sqrt{[a+1][b+1][c+1]}}{\hat{\Theta}(a,b,c)}},$$ showing that this factor can be taken to be positive real.
$\hfill \Box$
\bigbreak

In Figure 38 we show how this modification of the vertex affects the non-zero term of the orthogonality of trivalent
vertices (compare with Figure 32). We refer to this as the ``modified bubble identity." The coefficient in the modified bubble identity is
$$\sqrt{ \frac{\Delta_{b}\Delta_{c}}{\Delta_{a}} } = (-1)^{(b+c-a)/2} \sqrt{\frac{[b+1][c+1]}{[a+1]}}$$ 
where $(a,b,c)$ form an admissible triple. In particular $b+c-a$ is even and hence this factor can be taken to be real.
\bigbreak

We rewrite the recoupling formula in this new basis and emphasize 
that the recoupling coefficients can be seen (for fixed external labels $a,b,c,d$) as a matrix transforming the horizontal ``double-$Y$" basis
to a vertically
disposed double-$Y$ basis. In Figures 39, 40 and 41 we have shown the form of this transformation,using the matrix notation
$$M[a,b,c,d]_{ij}$$ for the modified recoupling coefficients. In Figure 39 we derive an explicit formula for these matrix elements. The proof of this 
formula follows directly from trivalent--vertex orthogonality (See Figures 32 and 35.), and is given in Figure 39. The result shown in Figure 39 and
Figure 40 is the  following formula for the recoupling matrix elements.
$$M[a,b,c,d]_{ij} = ModTet
\left( \begin{array}{ccc}
a &  b & i \\
c & d & j \\
\end{array} \right)/\sqrt{\Delta_{a}\Delta_{b}\Delta_{c}\Delta_{d}}$$
where $\sqrt{\Delta_{a}\Delta_{b}\Delta_{c}\Delta_{d}}$ is short-hand for the product
$$\sqrt{ \frac{\Delta_{a}\Delta_{b}}{\Delta_{j}} }\sqrt{ \frac{\Delta_{c}\Delta_{d}}{\Delta_{j}} } \Delta_{j}$$ 
$$= (-1)^{(a+b-j)/2}(-1)^{(c+d-j)/2} (-1)^{j} \sqrt{ \frac{[a+1][b+1]}{[j+1]}}\sqrt{ \frac{[c+1][d+1]}{[j+1]}} [j+1]$$
$$ = (-1)^{(a+b+c+d)/2}\sqrt{[a+1][b+1][c+1][d+1]}$$
In this form, since
$(a,b,j)$ and $(c,d,j)$ are admissible triples, we see that this coeffient can be taken to be real, and its value is
independent of the choice of $i$ and $j.$
The matrix $M[a,b,c,d]$ is real-valued.
\bigbreak

\noindent It follows from Figure 33 (turn the diagrams by ninety degrees) that 
$$M[a,b,c,d]^{-1} = M[b,d,a,c].$$
In Figure 42 we illustrate the formula
$$M[a,b,c,d]^{T} = M[b,d,a,c].$$ It follows from this formula that 
$$M[a,b,c,d]^{T} = M[a,b,c,d]^{-1}.$$ {\it Hence $M[a,b,c,d]$ is an orthogonal, real-valued matrix.}

$$ \picill3inby2in(F37)  $$
\begin{center}
{\bf Figure 37 - Modified Three Vertex}
\end{center}
\bigbreak

$$ \picill3inby3in(F38)  $$
\begin{center}
{\bf Figure 38 - Modified Bubble Identiy}
\end{center}
\bigbreak

$$ \picill3inby5in(F39)  $$
\begin{center}
{\bf Figure 39 - Derivation of Modified Recoupling Coefficients}
\end{center}
\bigbreak

$$ \picill3inby2.5in(F40)  $$
\begin{center}
{\bf Figure 40 - Modified Recoupling Formula}
\end{center}
\bigbreak

$$ \picill3inby2in(F41)  $$
\begin{center}
{\bf Figure 41 - Modified Recoupling Matrix}
\end{center}
\bigbreak

$$ \picill3inby3in(F42)  $$
\begin{center}
{\bf Figure 42 - Modified Matrix Transpose}
\end{center}
\bigbreak

\noindent {\bf Theorem 2.} {\it In the Temperley-Lieb theory we obtain unitary (in fact real orthogonal) recoupling transformations when the bracket
variable $A$ has the form $A = e^{i\pi/2r}$ for $r$ a positive integer. Thus we obtain families of unitary representations of the Artin braid group
from the recoupling  theory at these roots of unity.} 
\bigbreak

\noindent {\bf Proof.} The proof is given the discussion above. 
$\hfill \Box$
\bigbreak

In Section 9 we shall show explictly how these methods work in the case of the Fibonacci model where $A = e^{3i\pi/5}$.  
\bigbreak

\section {Fibonacci Particles}

In this section and the next we detail how the Fibonacci model for anyonic quantum computing \cite{Kitaev,Preskill} can be constructed by using a version of
the two-stranded  bracket polynomial and a generalization of Penrose spin networks. This is a fragment of the Temperly-Lieb recoupling theory \cite{KL}. We
already gave in the preceding sections a general discussion of the theory of spin networks and their relationship with quantum computing.
\bigbreak

The Fibonacci model is a $TQFT$ that is based on a single ``particle" with two states that we shall call the {\it marked state} and the 
{\it unmarked state}. The particle in the marked state can interact with itself either to produce a single particle in the marked state, or 
to produce a single particle in the unmarked state. The particle in the unmarked state has no influence in interactions (an unmarked state interacting
with any state $S$ yields that state $S$). 
One way to indicate these two interactions symbolically is to use a box,for the marked state and a blank space for the unmarked state.
Then one has two modes of interaction of a box with itself:
\begin{enumerate}
\item Adjacency: $\fbox{~} ~~ \fbox{~}$
\smallbreak
\noindent and 
\item Nesting: $\fbox{ \fbox{~~} }.$
\end{enumerate}

\noindent With this convention we take the adjacency interaction to yield a single box, and the nesting interaction to produce nothing:

$$\fbox{~} ~~ \fbox{~} = \fbox{~}$$
$$\fbox{ \fbox{~~} } =  $$

\noindent We take the notational opportunity to denote nothing by an asterisk (*). The syntatical rules for operating the asterisk are
Thus the asterisk is a stand-in for no mark at all and it can be erased or placed wherever it is convenient to do so.
Thus $$\fbox{ \fbox{~~} } = *. $$

$$ \picill3inby1.3in(F43)  $$

\begin{center}
{\bf Figure 43 - Fibonacci Particle Interaction}
\end{center}

We shall make a recoupling theory based on this particle, but it is worth noting some of its purely combinatorial properties first.
The arithmetic of combining boxes (standing for acts of distinction) according to these rules has been studied and formalized in 
\cite{LOF} and correlated with Boolean algebra and classical logic. Here {\em within} and {\em next to} are ways to refer to the two sides delineated by
the given distinction. From this point of view, there are two modes of relationship (adjacency and nesting) that arise at once in the presence of a
distinction.  
\bigbreak

$$ \picill3inby3.5in(F44)  $$

\begin{center}
{\bf Figure 44 - Fibonacci Trees}
\end{center}
\bigbreak

From here on we shall denote the Fibonacii particle by the letter $P.$
Thus the two possible interactions of $P$ with itself are as follows.
\begin{enumerate}
\item $P,P \longrightarrow *$
\item $P,P \longrightarrow P$
\end{enumerate}

\noindent In Figure 43 we indicate in small tree diagrams the two possible interactions of the particle  $P$ with itself.
In the first interaction the particle vanishes, producing the asterix. In the second interaction the particle
 a single copy of $P$ is produced. These are the two basic actions of a single distinction relative to itself, and they
constitute our formalism for this very elementary particle. 
\bigbreak

In Figure 44, we have indicated the different results of particle
processes where we begin with a left-associated tree structure with three branches, all marked and then four branches all marked.
In each case we demand that the particles interact successively to produce an unmarked particle in the end, at the root of the tree.
More generally one can consider a left-associated tree with $n$ upward branches and one root. Let $T(a_1,a_2, \cdots , a_n : b)$ denote such
a tree with particle labels $a_1, \cdots, a_n$ on the top and root label $b$ at the bottom of the tree. We consider all possible processes
(sequences of particle interactions) that start with the labels at the top of the tree, and end with the labels at the bottom of the tree.
Each such sequence is regarded as a basis vector in a complex vector space 
$$V^{a_1,a_2, \cdots , a_n}_{b}$$
associated with the tree. In the case where all the labels are marked at the top and the bottom label is unmarked, we shall denote this tree
by $$V^{111 \cdots 11}_{0} = V^{(n)}_{0}$$ where $n$ denotes the number of upward branches in the tree. We see from Figure 44 that the dimension 
of $V^{(3)}_{0}$ is $1,$ and that $$dim(V^{(4)}_{0}) = 2.$$ This means that $V^{(4)}_{0}$ is a natural candidate in this context for the two-qubit 
space.
\bigbreak

Given the tree $T(1,1,1,\cdots, 1:0)$ ($n$ marked states at the top, an unmarked state at the bottom), a process basis vector in $V^{(n)}_{0}$
is in direct correspondence with a string of boxes and asterisks ($1$'s and $0$'s) of length $n-2$ with no repeated asterisks and ending in a marked state.
See Figure 44 for an illustration of the simplest cases. It follows from this
that $$dim(V^{(n)}_{0}) = f_{n-2}$$ where $f_k$ denotes the $k$-th Fibonacci number:
$$f_0 = 1, f_1 = 1, f_2 = 2, f_3 = 3, f_4 = 5, f_5= 8, \cdots$$ where $$f_{n+2} = f_{n+1} + f_{n}.$$
The dimension formula for these spaces follows from the fact that there are $f_{n}$ sequences of length $n-1$ of marked and unmarked states with no
repetition of an unmarked state. This fact is illustrated in Figure 45.
\bigbreak  

$$ \picill3inby2.5in(F45)  $$

\begin{center}
{\bf Figure 45 - Fibonacci Sequence}
\end{center}
\bigbreak

\section{The Fibonacci Recoupling Model}
We now show how to make a model for recoupling the Fibonacci particle by using the Temperley Lieb recoupling theory and the bracket polynomial.
Everything we do in this section will be based on the 2-projector, its properties and evaluations based on the bracket polynomial model for the Jones
polynomial. While we have outlined the general recoupling theory based on the bracket polynomial in earlier sections of this paper,
the present section is  self-contained, using only basic information about the bracket polyonmial, and the essential properties of the 
2-projector as shown in Figure 46. In this figure we state the definition of the 2-projector, list its two main properties (the operator is idempotent and
a self-attached strand yields a zero evaluation) and give diagrammatic proofs of these properties.

$$ \picill3inby3in(F46)  $$

\begin{center}
{\bf Figure 46 - The 2-Projector}
\end{center}
\bigbreak

In Figure 47, we show the essence of the Temperley-Lieb recoupling model for the Fibonacci particle. The Fibonaccie particle is, in this mathematical
model, identified with the 2-projector itself. As the reader can see from Figure 47, there are two basic interactions of the 2-projector with itself, 
one giving a 2-projector, the other giving nothing. This is the pattern of self-iteraction of the Fibonacci particle. There is a third possibility,
depicted in Figure 47, where two 2-projectors interact to produce a 4-projector. We could remark at the outset, that the 4-projector will be zero if we
choose the bracket polynomial variable $A = e^{3 \pi/5}.$ Rather than start there, we will assume that the 4-projector is forbidden and deduce (below)
that the theory has to be at this root of unity.

$$ \picill3inby3in(F47)  $$

\begin{center}
{\bf Figure 47 - Fibonacci Particle as 2-Projector}
\end{center}
\bigbreak

\noindent Note that in Figure 47 we have adopted a single strand notation for the particle interactions, with a solid strand corresponding to the
marked particle, a dotted strand (or nothing) corresponding to the unmarked particle.  A dark vertex indicates either an interaction point, or it
may be used to indicate the single strand is shorthand for two ordinary strands. Remember that these are all shorthand expressions for underlying
bracket polynomial calculations.
\bigbreak

In Figures 48, 49, 50, 51, 52 and 53 we have provided complete diagrammatic calculations of all of the relevant small nets and evaluations that 
are useful in the two-strand theory that is being used here. The reader may wish to skip directly to Figure 54 where we determine the 
form of the recoupling coefficients for this theory. We will discuss the resulting algebra below.
\bigbreak

For the reader who does not want to skip the next collection of Figures, here is a guided tour. Figure 48 illustrates three three basic nets in case of two 
strands. These are the theta, delta and tetrahedron nets. In this Figure we have shown the decomposition on the theta and delta nets in terms of
2-projectors. The Tetrahedron net will be similarly decomposed in Figures 52 and 53. The theta net is denoted $\Theta,$ the delta by $\Delta,$ and the
tetrahedron by $T.$  In Figure 49 we illustrate how a pedant loop has a zero evaluation. In Figure 50 we use the identity in Figure 49 to show how
an interior loop (formed by two trivalent vertices) can be removed and replaced by a factor of $\Theta/\Delta.$ Note how, in this figure, line two proves
that one network is a multiple of the other, while line three determines the value of the multiple by closing both nets.
\bigbreak

\noindent Figure 51 illustrates the explicit calculation of the delta and theta nets. The figure begins with a calculation of the result of closing
a single strand of the 2-projector. The result is a single stand multiplied by $(\delta - 1/\delta)$ where $\delta = -A^2 - A^{-2},$ and $A$ is the bracket
polynomial parameter. We then find that $$\Delta = \delta^{2} - 1$$ and 
$$\Theta = (\delta - 1/\delta)^{2} \delta - \Delta/\delta = (\delta -1/\delta)(\delta^{2} - 2).$$  
\bigbreak

\noindent Figures 52 and 53 illustrate the calculation of the value of the tetrahedral network $T.$ The reader should note the first line of 
Figure 52 where the tetradedral net is translated into a pattern of 2-projectors, and simplified. The rest of these two figures are a diagrammatic
calculation, using the expansion formula for the 2-projector. At the end of Figure 53 we obtain the formula for the tetrahedron
$$T = (\delta - 1/\delta)^{2}(\delta^{2} - 2) - 2\Theta/\delta.$$ 

$$ \picill3inby3in(F48)  $$

\begin{center}
{\bf Figure 48 - Theta, Delta and Tetrahedron}
\end{center}
\bigbreak

$$ \picill3inby3in(F49)  $$

\begin{center}
{\bf Figure 49 - LoopEvaluation--1}
\end{center}
\bigbreak

$$ \picill3inby3.5in(F50)  $$

\begin{center}
{\bf Figure 50 - LoopEvaluation--2}
\end{center}
\bigbreak

$$ \picill3inby4in(F51)  $$

\begin{center}
{\bf Figure 51 - Calculate Theta, Delta}
\end{center}
\bigbreak

$$ \picill3inby2.7in(F52)  $$

\begin{center}
{\bf Figure 52 - Calculate Tetrahedron -- 1}
\end{center}
\bigbreak

$$ \picill3inby2.5in(F53)  $$

\begin{center}
{\bf Figure 53 - Calculate Tetrahedron -- 2}
\end{center}
\bigbreak

Figure 54 is the key calculation for this model. In this figure we assume that the recoupling formulas involve only $0$ and $2$ strands, with 
$0$ corresponding to the null particle and $2$ corresponding to the 2-projector. ($2 + 2 = 4$ is forbidden as in Figure 47.) From this assumption we
calculate that the recoupling matrix is given by 
$$ F = 
\left( \begin{array}{cc}
a & b \\
c & d \\
\end{array} \right) =
\left( \begin{array}{cc}
1/\Delta & \Delta/\Theta \\
\Theta/\Delta^{2} & T \Delta/\Theta^{2} \\
\end{array} \right)
$$

$$ \picill5inby5in(F54)  $$

\begin{center}
{\bf Figure 54 - Recoupling for 2-Projectors}
\end{center}
\bigbreak

$$ \picill5inby6.5in(F55)  $$

\begin{center}
{\bf Figure 55 - Braiding at the Three-Vertex}
\end{center}
\bigbreak

$$ \picill6inby6.5in(F56)  $$

\begin{center}
{\bf Figure 56 - Braiding at the Null-Three-Vertex}
\end{center}
\bigbreak

\noindent Figures 55 and 56 work out the exact formulas for the braiding at a three-vertex in this theory. When the 3-vertex has three marked lines,
then the braiding operator is multiplication by $-A^{4},$ as in Figure 55. When the 3-vertex has two marked lines, then the braiding operator is
multiplication by $A^{8},$ as shown in Figure 56.
\bigbreak

\noindent Notice that it follows from the symmetry of the diagrammatic recoupling formulas of Figure 54 that 
{\it the square of the recoupling matrix $F$ is equal to 
the identity.} That is, 
$$\left( \begin{array}{cc}
1 & 0 \\
0 & 1 \\
\end{array} \right) = F^{2} =
\left( \begin{array}{cc}
1/\Delta & \Delta/\Theta \\
\Theta/\Delta^{2} & T \Delta/\Theta^{2} \\
\end{array} \right)
\left( \begin{array}{cc}
1/\Delta & \Delta/\Theta \\
\Theta/\Delta^{2} & T \Delta/\Theta^{2} \\
\end{array} \right) =$$ 
$$\left( \begin{array}{cc}
 1/\Delta^{2} + 1/\Delta & 1/\Theta + T\Delta^{2}/\Theta^{3} \\
\Theta/\Delta^{3} + T/(\Delta\Theta) & 1/\Delta + \Delta^{2} T^{2}/\Theta^{4} \\
\end{array} \right).$$
Thus we need the relation
$$1/\Delta + 1/\Delta^{2} = 1.$$
This is equivalent to saying that 
$$\Delta^{2} = 1 + \Delta,$$ a quadratic equation whose solutions are
$$\Delta = (1 \pm \sqrt{5})/2.$$
Furthermore, we know that $$\Delta = \delta^{2} - 1$$ from Figure 51.
Hence $$\Delta^{2} = \Delta + 1 = \delta^{2}.$$
We shall now specialize to the case where
$$\Delta = \delta = (1 + \sqrt{5})/2,$$
leaving the other cases for the exploration of the reader.
We then take $$A = e^{3\pi i/5}$$ so that 
$$\delta = -A^{2} - A^{-2} = -2cos(6\pi/5) = (1 + \sqrt{5})/2.$$
\bigbreak

Note that $\delta - 1/\delta = 1.$ Thus
$$\Theta = (\delta - 1/\delta)^{2} \delta - \Delta/\delta = \delta - 1.$$
and
$$T = (\delta - 1/\delta)^{2}(\delta^{2} - 2) - 2\Theta/\delta = (\delta^{2} - 2) - 2(\delta - 1)/\delta$$
$$= (\delta - 1)(\delta -2)/\delta = 3\delta - 5.$$
Note that $$T = -\Theta^{2}/\Delta^{2},$$ from which it follows immediately that 
$$F^{2} = I.$$ This proves that we can satisfy this model when $\Delta = \delta = (1 + \sqrt{5})/2.$
\bigbreak

\noindent For this specialization we see that the matrix $F$ becomes
$$ F = 
\left( \begin{array}{cc}
1/\Delta & \Delta/\Theta \\
\Theta/\Delta^{2} & T \Delta/\Theta^{2} \\
\end{array} \right) =
\left( \begin{array}{cc}
1/\Delta & \Delta/\Theta \\
\Theta/\Delta^{2} & (-\Theta^{2}/\Delta^{2}) \Delta/\Theta^{2} \\
\end{array} \right) =
\left( \begin{array}{cc}
1/\Delta & \Delta/\Theta \\
\Theta/\Delta^{2} & -1/\Delta \\
\end{array} \right)$$
This version of $F$ has square equal to the identity  independent of the value of $\Theta,$ so long as $\Delta^{2} = \Delta + 1.$
\bigbreak

\noindent {\bf The Final Adjustment.} Our last version of $F$ suffers from a lack of symmetry. It is not a symmetric matrix, and hence
not unitary. A final adjustment of the model gives this desired symmetry. {\it Consider the result of replacing each trivalent vertex (with three 2-projector
strands) by a multiple by a given quantity $\alpha.$} Since the $\Theta$ has two vertices, it will be multiplied by $\alpha^{2}.$ Similarly,
the tetradhedron $T$ will be multiplied by $\alpha^{4}.$ The $\Delta$ and the $\delta$ will be unchanged. Other properties of the model will remain 
unchanged. The new recoupling matrix, after such an adjustment is made, becomes
$$\left( \begin{array}{cc}
1/\Delta & \Delta/\alpha^{2}\Theta \\
\alpha^{2}\Theta/\Delta^{2} & -1/\Delta \\
\end{array} \right)$$
For symmetry we require $$\Delta/(\alpha^{2}\Theta) = \alpha^{2}\Theta/\Delta^{2}.$$ We take $$\alpha^{2} = \sqrt{\Delta^{3}}/\Theta.$$
With this choice of $\alpha$ we have $$\Delta/(\alpha^{2}\Theta) = \Delta \Theta/(\Theta \sqrt{\Delta^{3}}) = 1/\sqrt{\Delta}.$$
Hence the new symmetric $F$ is given by the equation
$$F =
\left( \begin{array}{cc}
1/\Delta & 1/\sqrt{\Delta} \\
1/\sqrt{\Delta} & -1/\Delta \\
\end{array} \right) =
\left( \begin{array}{cc}
\tau & \sqrt{\tau} \\
\sqrt{\tau} & -\tau \\
\end{array} \right)$$
where $\Delta$ is the golden ratio and $\tau = 1/\Delta$.
This gives the Fibonacci model. Using Figures 55 and 56, we have that the local braiding matrix for the model is given by the formula
below with $A = e^{3\pi i/5}.$
$$R = 
\left( \begin{array}{cc}
-A^{4} & 0 \\
0 & A^{8} \\
\end{array} \right)=
\left( \begin{array}{cc}
e^{4\pi i/5} & 0 \\
0 & -e^{2\pi i/5} \\
\end{array} \right).$$
\bigbreak

The simplest example of a braid group representation arising from this theory is the representation of the three strand braid group generated by
$S_{1}= R$ and $S_{2} = FRF$ (Remember that $F=F^{T} = F^{-1}.$). The matrices $S_{1}$ and $S_{2}$ are both unitary, and they generate a dense subset of
the unitary group $U(2),$ supplying the first part of the transformations needed for quantum computing.
\bigbreak

\section{Quantum Computation of Colored Jones Polynomials and the Witten-Reshetikhin-Turaev Invariant}
In this section we make some brief comments on the quantum computation of colored Jones polynomials. This material will be expanded in a subsequent
publication.

$$ \picill3inby6in(F57)  $$

\begin{center}
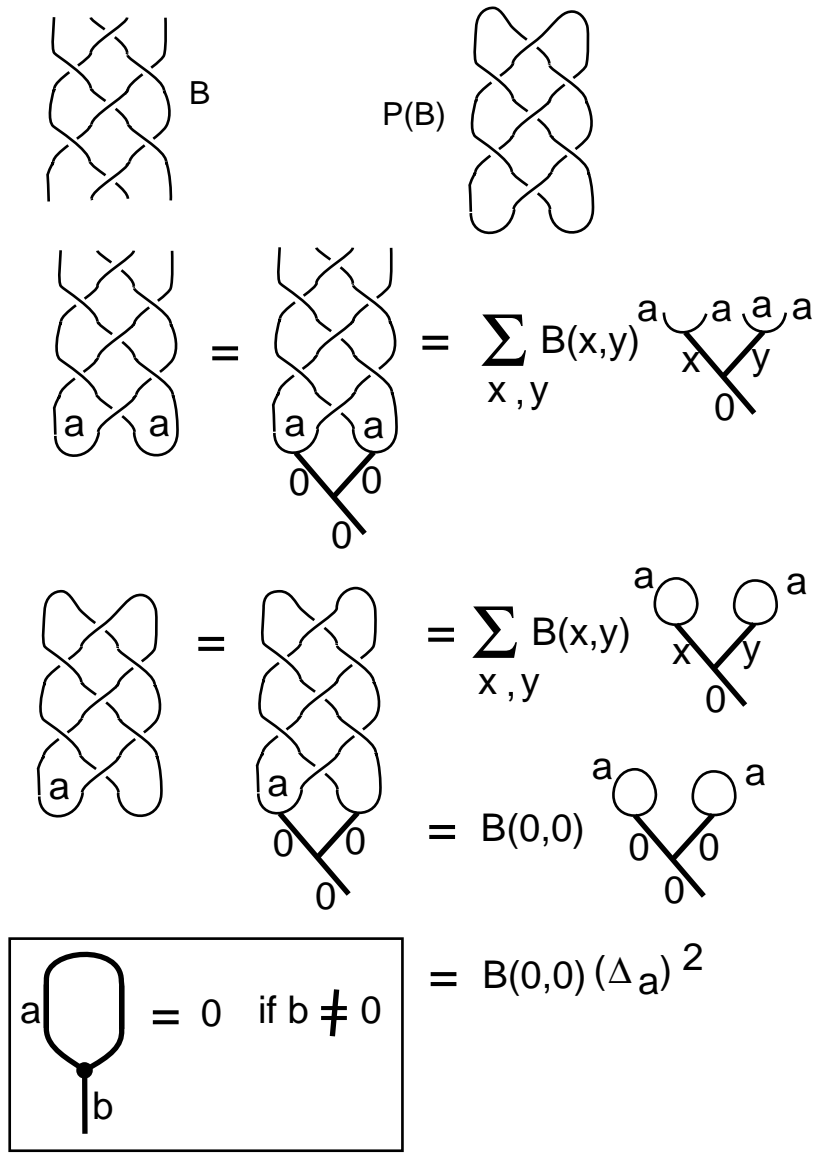
{\bf Figure 57 - Evaluation of the Plat Closure of a Braid}
\end{center}
\bigbreak

First, consider Figure 57. In that figure we illustrate the calculation of the evalutation of the {\it ($a$) - colored bracket polynomial} for
the {\it plat closure} $P(B)$  of a braid $B$. 
The reader can infer the definition of the plat closure from Figure 57. One takes a braid on an even number of strands and closes the top strands with
each other in a row of maxima. Similarly, the bottom strands are closed with a row of minima. It is not hard to see that any knot or link can be represented
as the plat closure  of some braid.
\bigbreak

$$ \picill5inby4in(F58)  $$

\begin{center}
{\bf Figure 58 - Dubrovnik Polynomial Specialization at Two Strands  }
\end{center}
\bigbreak

The ($a$) - colored bracket polynonmial of a link $L$, denoted $<L>_{a},$ is the evaluation of that link where each single strand has been replaced by $a$
parallel strands and the insertion of Jones-Wenzl projector (as discussed in Section 7). We then see that we can use our discussion of the Temperley-Lieb
recoupling theory as in sections 7,8 and 9 to compute the value of the colored bracket polynomial for the plat closure $PB.$ As shown in Figure 57, we regard
the braid as  acting on a process space $V^{a,a,\cdots,a}_{0}$ and take the case of the action on the vector $v$ whose process space coordinates are all
zero. Then the action of the braid takes the form $$Bv(0,\cdots,0) = \Sigma_{x_{1},\cdots,x_{n}} B(x_{1},\cdots,x_{n}) v(x_{1},\cdots,x_{n})$$
where $B(x_{1},\cdots,x_{n})$ denotes the matrix entries for this recoupling transformation and $v(x_{1},\cdots,x_{n})$ runs over a basis for the 
space $V^{a,a,\cdots,a}_{0}.$ Here $n$ is even and equal to the number of braid strands. In the figure we illustrate with $n=4.$ Then, as the figure shows,
when we close the top of the braid action to form $PB,$ we cut the sum down to the evaluation of just one term. In the general case we will get
$$<PB>_{a} = B(0,\cdots,0)\Delta_{a}^{n/2}.$$ The calculation simplifies to this degree because of the vanishing of loops in the recoupling graphs.
The vanishing result is stated in Figure 57, and it is proved in the case $a =2$ in Figure 49.
\bigbreak

The {\it colored Jones polynomials} are normalized versions of the colored bracket polymomials, differing just by a normalization factor.
\bigbreak

In order to consider quantumn computation of the colored bracket or colored Jones polynomials, we therefore can consider quantum computation of the 
matrix entries $B(0,\cdots,0).$ These matrix entries in the case of the roots of unity $A = e^{i\pi/2r}$ and for the $a=2$ Fibonacci model with 
$A= e^{3i\pi/5}$ are parts of the diagonal entries of the unitary transformation that represents the braid group on the process space  $V^{a,a,\cdots,a}_{0}.$
{\it We can obtain these  matrix entries by using the Hadamard test as described in section 4.} As a result we get relatively efficient quantum
algoritms for the colored Jones polynonmials at these roots of unity, in essentially the same framework as we described in section 4, but for braids of 
arbitrary size. The computational complexity of these models is essentially the same as the models for the Jones polynomial discussed in \cite{Ah1}.
We reserve discussion of these issues to a subsequent publication.
\bigbreak

It is worth remarking here that these algorithms give not only quantum algorithms for computing the colored bracket and Jones polynomials, but also for
computing the Witten-Reshetikhin-Turaev ($WRT$) invariants at the above roots of unity. The reason for this is that the $WRT$ invariant, in unnormalized
form is given as a finite sum of colored bracket polynomials:
$$WRT(L) = \Sigma_{a = 0}^{r-2} \Delta_{a} <L>_{a},$$
and so the same computation as shown in Figure 57 applies to the $WRT.$ This means that we have, in principle, a quantum algorithm for the computation of the
Witten functional integral \cite{Witten} via this knot-theoretic combinatorial topology. It would be very interesting to understand a more direct approach to
such a computation via quantum field theory and functional integration.
\bigbreak

Finally, we note that in the case of the Fibonacci model, the ($2$)-colored bracket polynomial is a special case of the Dubrovnik version of the 
Kauffman polynomial \cite{IRI}. See Figure 58 for diagammatics that resolve this fact. The skein relation for the Dubrovnik polynomial is boxed in this
figure. Above the box, we show how the double strands with projectors reproduce this relation. This observation means that in the Fibonacci model, the
natural underlying knot polynomial is a special evaluation of the Dubrovnik polynomial, and the Fibonacci model  can be used to perform quantum computation
for the values of this invariant.

\end{document}